\documentclass{article}

\usepackage[centering]{geometry}
\usepackage{amsfonts}
\usepackage{amsmath}
\usepackage[applemac]{inputenc}
\usepackage[english]{babel}
\usepackage{color}
\usepackage{hyperref}
\usepackage{subcaption}

\usepackage{amssymb}
\usepackage{amsopn}
\usepackage{mathrsfs}

\usepackage{mathtools}
\mathtoolsset{showonlyrefs}

\usepackage[textwidth=2.1cm]{todonotes}

\newcommand{\be}{\begin{equation}}
\newcommand{\ee}{\end{equation}}

\usepackage[normalem]{ulem}

\usepackage{todonotes}
\allowdisplaybreaks

\title{A neuro-mathematical model \\ for size and context related illusions.}
\author{B. Franceschiello, A. Sarti, G. Citti}
\date{March 2019}

\begin{document}

\maketitle

\begin{abstract}
We provide here a mathematical model of size/context illusions, inspired by the functional architecture of the visual cortex. We first recall previous models of scale and orientation, in particular  \cite{sarti2008symplectic}, and simplify it, only considering the feature of scale. Then we recall the deformation model of illusion, introduced by  \cite{franceschiello2017neuro}  
to describe orientation related GOIs, and adapt it to size illusion. We finally apply the model to the Ebbinghaus and Delboeuf illusions, 
validating the results by comparing with experimental data from 
\cite{massaro1971judgmental} and \cite{roberts2005roles}.
\end{abstract}

\section{Introduction}
\label{intro}
Geometrical-optical illusions (GOIs) are a class of phenomena first discovered by German physicists and physiologists in the late XIX century, among them Oppel and Hering (\cite{oppel1855uber}, \cite{Her_1}), and can be defined as situations where a perceptual mismatch between the visual stimulus and its geometrical properties arise \cite{westheimer2008illusions}. Those illusions are typically analyzed according to the main geometrical features of the stimulus, whether it is contours orientation, contrast, context influence, size or a combination of the above mentioned ones (\cite{westheimer2008illusions,ninio2014geometrical,eagleman2001visual}).
\begin{figure}[htbp]
\centering
\includegraphics[width = 0.35\textwidth]{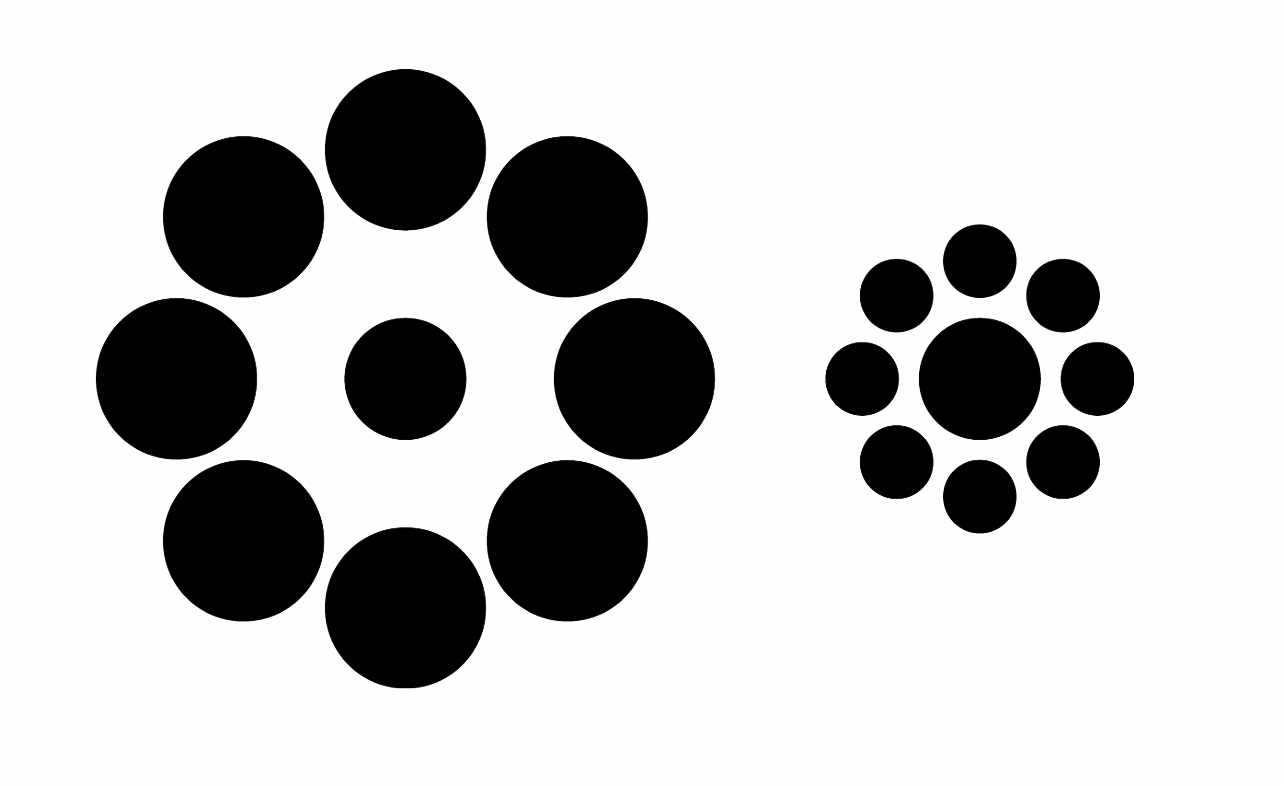}\;\;\;\;\;\;
\includegraphics[width = 0.35\textwidth]{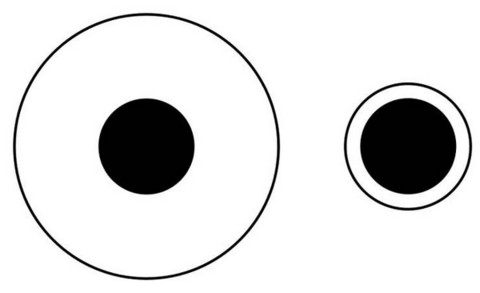}
\caption{The Ebbinghaus illusion (left) and the Delboeuf illusion (right)  }
\label{ebb1}
\end{figure}

In this work we are mainly interested in size and context related phenomena, a class of stimuli where the size of the surroundings elements induces a misperception of the central target width. In figure \ref{ebb1}, two famous effects are presented, the Ebbinghaus and Delboeuf illusions: the presence of circular inducers (figure \ref{ebb1}, left) and of an annulus (figure \ref{ebb1}, right) varies the perceived sizes of the central targets. These phenomena have been named after their discoverers, the German psychologist Hermann Ebbinghaus (1850 - 1909), and the Belgian philosopher and mathematician Joseph Remi Leopold Delboeuf (1831 - 1896), \cite{delboeuf1865note}. The Ebbinghaus phenomenon has been popularized in the English- speaking world by Edward B. Titchener in a 1901 textbook of experimental psychology, and this is the reason why it is also called Titchener illusion \cite{roberts2005roles}.

\begin{figure}
\centering
\includegraphics[width=0.4\textwidth]{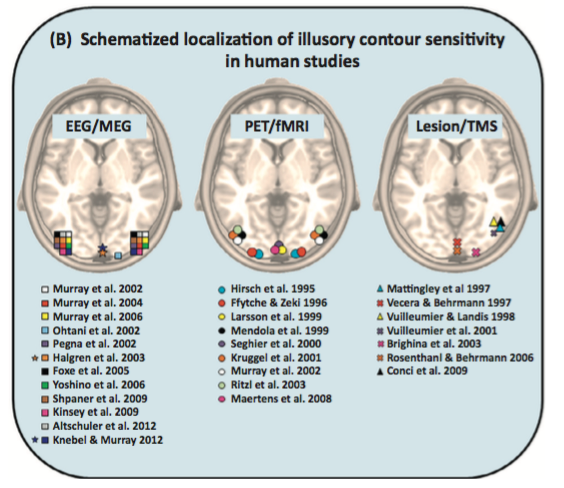} 
\caption{Quoting from Murray and Herrmann \cite{murray2013illusory}: (B) Schematic localization of Illusory contours (IC) sensitivity in human studies. The colored symbols indicate the approximate locations of IC sensitivity for human studies using electroencephalography (EEG)/magnetoencephalography (MEG) source estimations (left), positron emission tomography (PET) and functional magnetic resonance imaging (fMRI) (middle), and lesion studies or transcranial magnetic stimulation (TMS) (right). The stars in the left panel indicate secondary and subsequent effects.}
\label{fig:3}
\end{figure}

The importance of studying these phenomena at a psychophysical and neuroimaging level lies in the fact that these phenomena provide insights about the functionality of the visual system (\cite{eagleman2001visual}).
Many studies show that at least  neurons the visual areas V1 and V2, carry signals related to illusory contours, and that signals in V2 are more robust than in V1 (\cite{von1984illusory, murray2002spatiotemporal}, reviews \cite{eagleman2001visual,murray2013illusory}), see figure \ref{fig:3} from \cite{murray2013illusory}. As for what concerns size and context-dependent phenomena such as those presented in figure \ref{ebb1}, it is not new that attention plays a huge role in modulating the visual response \cite{yan2014attention}. A further proof for that lies in the usage of these context dependent illusions for proving different perceptual mechanisms, related to attention, in cross-cultural study, see \cite{doherty2008context,bremner2016effects,fonteneau2008cultural}. 

Geometrical models for optical illusions related to orientation perception were proposed by the pioneering work of Hoffman \cite{hoffman1971visual}, in term of Lie groups, and then by Smith, \cite{smith1978descriptive}, who stated that the apparent curve of GOIs (where the main feature questioned during the perceptual processing is orientation) can be modeled by a first-order differential equation. A first attempt was performed also by Walker (\cite{walker1973mathematical}), who tried to combine neural theory of receptive field excitation together with mathematical tools to explain misperception of straight lines in GOIs.
These results, together with \cite{smith1978descriptive} and Ehm and Wackerman in~\cite{ehm2012modeling}, introduced a quantitative analysis of the perceived distortion. On the other hand another possible way to go is to use a Bayesian approach to model the neural activity, approach who lies onto Helmholtz's theory \cite{von2005treatise,geisler2002illusions}. These methods allow to consider how prior experience influences perception, \cite{knill1996perception}, and were applied to motion illusions by Weiss et al. in \cite{weiss2002motion}. Ferm\"{u}ller and Malm in \cite{Ferm} attributed the perception of geometric optical illusions to the statistics of visual computations. 
 More recently the authors in \cite{franceschiello2017mathematical, franceschiello2017neuro} proposed a model for orientation based geometrical illusions inspired by the functionality of simple cells of the visual cortex \cite{citti2006cortical}. 
Geometric models of the functionality if the visual cortex were proposed by Hoffman  \cite{hoffman1989visual}, Mumford in \cite{mumford1994elastica}, Williams and Jacobs in \cite{williams1997stochastic}, and more recently by  Petitot and Tondut in \cite{petitot1999naturalizing} and  Citti and Sarti in \cite{citti2006cortical}. These models were main focused on orientation selectivity, but they have also been extended to describe  scale selectivity in    \cite{sarti2009functional}  \cite{sarti2008symplectic}.

The aim of this paper is therefore to extend the work in \cite{franceschiello2017mathematical, franceschiello2017neuro}, to illusion of size and scale, starting from the cortical model of Sarti, Citti and Petitot (\cite{sarti2009functional}). In their models families of simple cells are characterized by a cortical connectivity and a functional geometry. The main idea is that the context modulates the connectivity metric and induces a deformation of the space, from which it will be possible to compute the displacement and the corresponding perceived misperception. An isotropic functional connectivity depending on the detected scale and on the distance between the objects composing the stimulus will be considered. It will follow an explanation concerning the implementation of the phenomena and a description of the numerical simulations performed to compute the perceived deformation. The computations will be in agreement with a judgemental study of Massaro et al. \cite{massaro1971judgmental}, as well as with the observations of how illusions change varying the distance between target and inducers, \cite{roberts2005roles}. To our knowledge, this is the first original contribution providing an interpretation to size related geometrical optical illusions.  

\section{Neurogeometry of the primary visual cortex and GOIs}
\label{sec:2}
Neuromathematical models target features encoding during early stages of the early visual process. The first geometric models of the functionality of the visual cortex
date back to the papers of Hoffmann \cite{hoffman1989visual} and Koenderink-van Doorn \cite{koenderink1987representation}. Citti and Sarti developed in \cite{citti2006cortical, sarti2008symplectic, sarti2015constitution, sarti2015constitution}, a theory of invariant perception in Lie groups, taking into account (separately or together)  different features: brightness orientation, scale, curvature, movement. Other papers applying instruments of Lie groups and differential geometry for the description of the visual cortex have been introduced by August and Zucker \cite{august2000curve}, Petitot and Tondut \cite{petitot1999vers}, Duits and Franken \cite{duits2010left,duits2010left2}.

\subsection{The receptive field of a cortical neuron}
\label{sec:2.1}
 
 \begin{figure}
 \centering
\includegraphics[width = 
 .9\columnwidth]{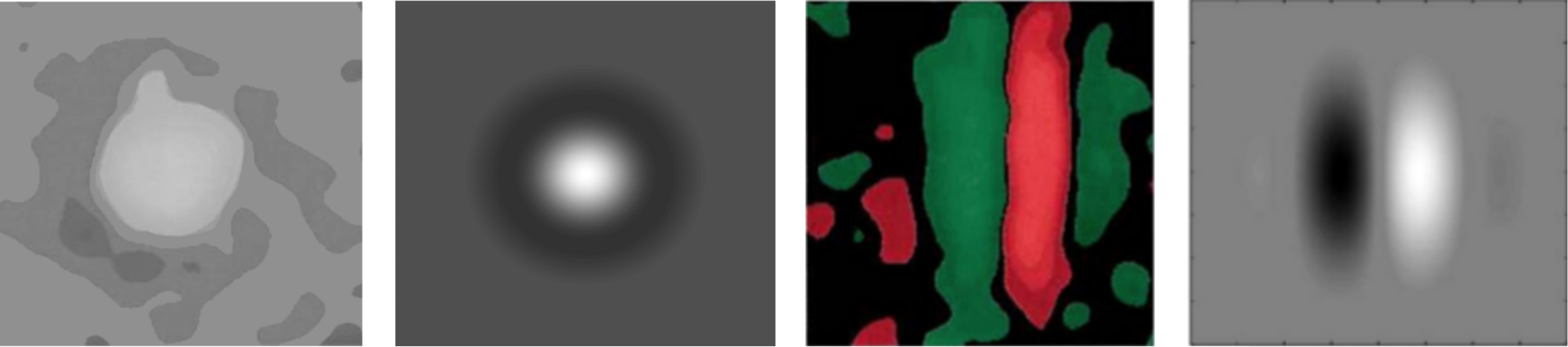}
 \caption{From left to right: in vivo registered radial symmetric receptive fields, see De Angelis et al. in \cite{deangelis1995receptive}; their model as Laplacian of a Gaussian; in vivo recorded odd receptive field (from De Angelis et al. in \cite{deangelis1995receptive}); their model as Gabor filter, see \eqref{receptive_profile}}
\label{fig:2e}
\end{figure}
The visual process is the result of several retinal and cortical mechanisms acting on the visual signal. The retina is the first part of the visual system responsible for the transmission of the signal, which passes through the Lateral Geniculate Nucleus, where a first preprocessing is performed, and arrives in the visual cortex, where it is further processed. 
The receptive field (RF) of a cortical neuron is the portion of the retina which the neuron reacts to, and the receptive profile (RP)  $\psi(\xi)$ is the function that models the activation of a cortical neuron when a stimulus is applied to a point $\xi = (\xi_1,\xi_2)$ of the retinal plane.
As an example, we recall that the RP of simple cells sensible of scale and orientation have been experimentally described by De Angelis in \cite{deangelis1995receptive}, and modelled as a Gabor filter in Daugman \cite{Daug}, Jones and Palmer \cite{jones1987evaluation}, see Figure \ref{fig:2e}.
If  $T_{x_1,x_2}$ is a translation of vector $(x_1,x_2)$,  $D_\sigma$ a dilation of amplitude 
$\sigma$ and $R_\theta$ is a rotation of an angle $\theta$, a good expression for the Gabor filters sensible to position $(x_1, x_2)$,  orientation and scale $(\theta, \sigma)$, is:
\begin{equation}
\psi_{x_1, x_2, \sigma, \theta} (\xi) = D_\sigma R_\theta\psi_0 T_{x_1, x_2}(\xi), \quad {\text{where }}\quad
\psi_0 (\xi) = \frac{1}{4 \pi} e^{-\frac{4\xi_1^2 + \xi_2^2}{8}}e^{2i\bar{b}\xi_2}. 
 \label{receptive_profile}
 \end{equation}

\subsection{Output of receptive profiles} 
\label{212}     

\begin{figure} 
\centering
\includegraphics[width=0.4\columnwidth]{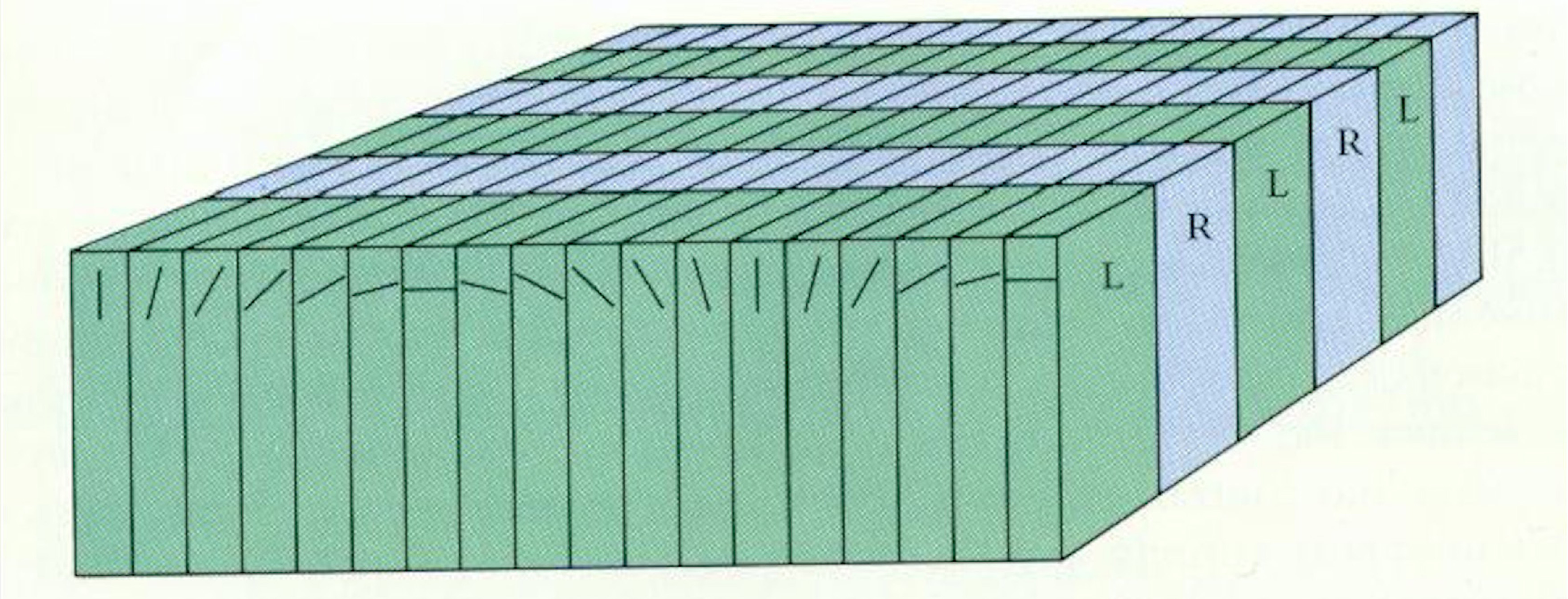}  \;\;\; \includegraphics[width=0.3\columnwidth]{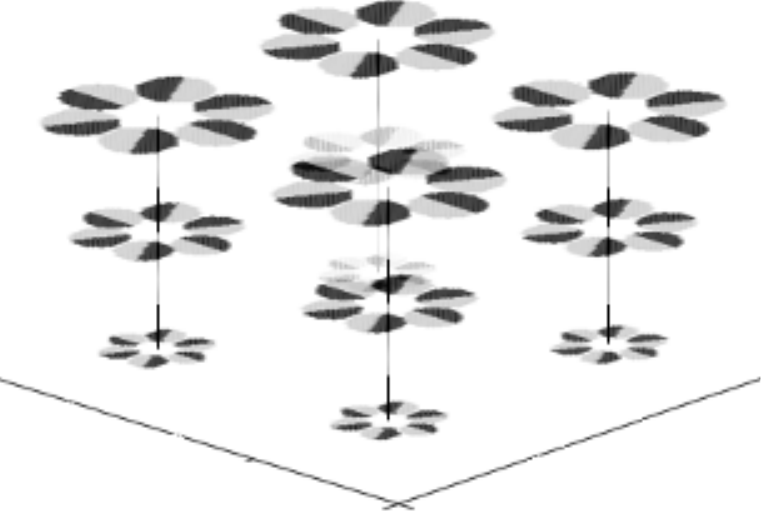}
\caption{Left: representation of the hypercolumnar structure, for the orientation parameter, where L and R represent the ocular dominance columns (Petitot \cite{petitot2008neurogeometrie}). Right: for each retinal position $(x_1,x_2)$, according to the model in \cite{sarti2009functional}, there is the set of all possible orientations and scale.}\label{hyper}
\end{figure}

Due to the retinotopic structure, there is an isomorphism between the retinal and cortical plane in V1, which we will discard in first approximation. Furthermore the hypercolumnar structure, discovered by the neurophysiologists Hubel and Wiesel in the 60s (\cite{hubel1977ferrier}), organizes the cells of V1/V2 in columns (called hypercolumns), each one covering a small part of the visual field $M \subset \mathbb{R}^2$ and corresponding to parameters such as orientation, scale, direction of movement, color, for a fixed retinal position $(x_1,x_2)$. Over each retinal point we will consider a whole hypercolumn of cells, each one sensitive to a specific instance of the considered feature $f$, see Figure \ref{hyper}.
We will then identify cells in the cortex by the three parameters $(x_1,x_2,f)$, where $(x_1,x_2)$ represents the position of the point and $f$ is a vector of extracted features. We will denote with $F$ the set of features, and consequently the cortical space will be identified as $R^2 \times F$.  

The retinal plane is identified with the $\mathbb{R}^2$-plane, whose local coordinates will be denoted with $x=(x_1,x_2)$. When a visual stimulus $I$ of intensity $I(x_1,x_2) : M \subset \mathbb{R}^2 \rightarrow \mathbb{R}^+$ activates the retinal layer of photoreceptors, the neurons whose RFs intersect $M$ spike and their spike frequencies $O(x_1,x_2,f)$ can be modeled (taking into account just linear contributions) as the integral of the signal $I(x_1,x_2)$ with the set of Gabor filters. The expression for this output is: 
\begin{equation}
O(x_1,x_2,f) = \int_M \, I(\xi_1, \xi_2)\, \psi_{(x_1,x_2,f)}\, (\xi_1, \xi_2) \, d\xi_1 d\xi_2 .
\label{output_simp_cel}
\end{equation}

\subsection{Cortical connectivity}
\label{sec:cc}
Note that the output is a higher dimensional function, defined on the cortical space. 
The lateral connectivity propagates this output in the cortical space $R^2 \times F$ giving rise to the cortical activity. 
This corresponds to spatial activity in the cortex, where neurons are parametrized by the variables $(x,f)$.
Interactions between synaptically coupled neurons occur via events called action potentials. A single action potential evokes a voltage change (post synaptic potential, PSP) in the postsynaptic element. Cortical connectivity has been measured in many families of cells, and it is strongly anisotropic. It has been proved that there is a relation between the shape of the receptive profiles, their connectivity and their functionality. 
A good model for the cortical connectivity can be obtained describing $R^2 \times F$ as a Lie group, endowed with a sub-Riemannian metric, or a symplectic structure, and choosing $K_F$ as a decreasing exponential function of the distance. Typically 
a fundamental solution of a  Fokker Planck equation, left invariant with respect to the group law has this property. 
Since we are interested into scale models, we recall here the model of Sarti, Citti and Petitot in \cite{sarti2008symplectic,petitot2008neurogeometrie} who proposed a model of scale and orientation selectivity in  $R^2 \times S^1 \times R^+$, with  the group low of translation, rotation and dilation. A basis of 
left invariant vector fields can be defined as:
$$X_1= \sigma(\cos\theta \partial_{x_1} +\sin\theta \partial_{x_2});\;  X_2=\sigma \partial_\theta;\; X_3=\sigma^2(-\sin\theta \partial_{x_1} +\cos\theta \partial_{x_2});\; X_4=-\sigma^2\partial_\sigma.$$
 \begin{figure}
\centering
\includegraphics[width=0.6\textwidth]{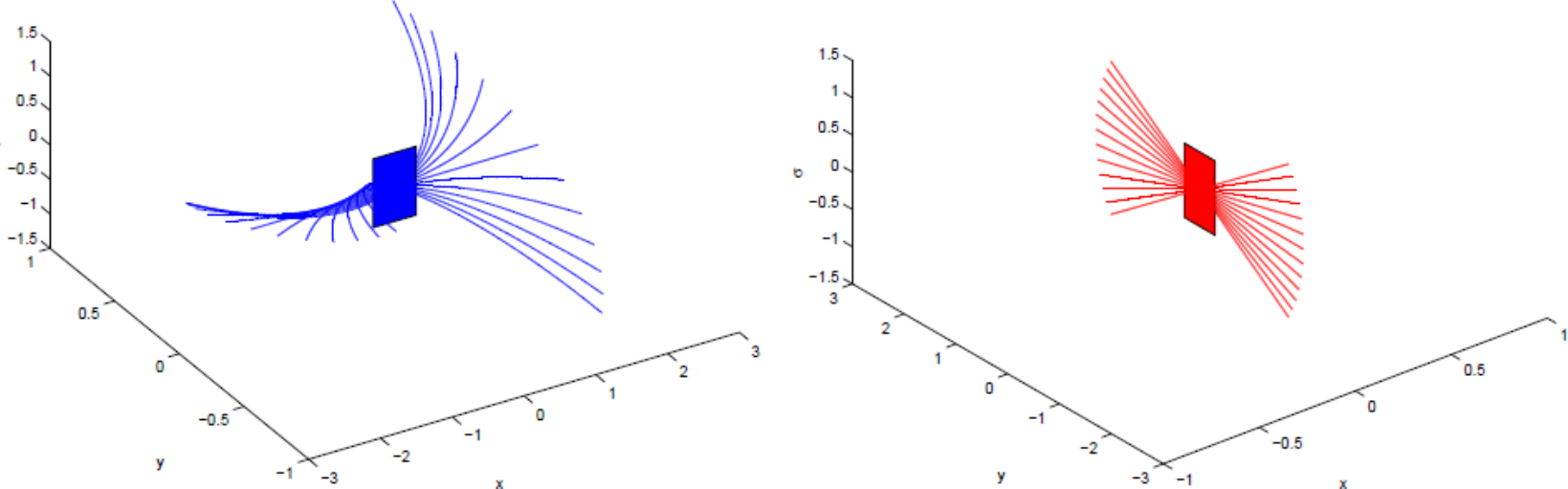}\\
\includegraphics[width=0.6\textwidth]{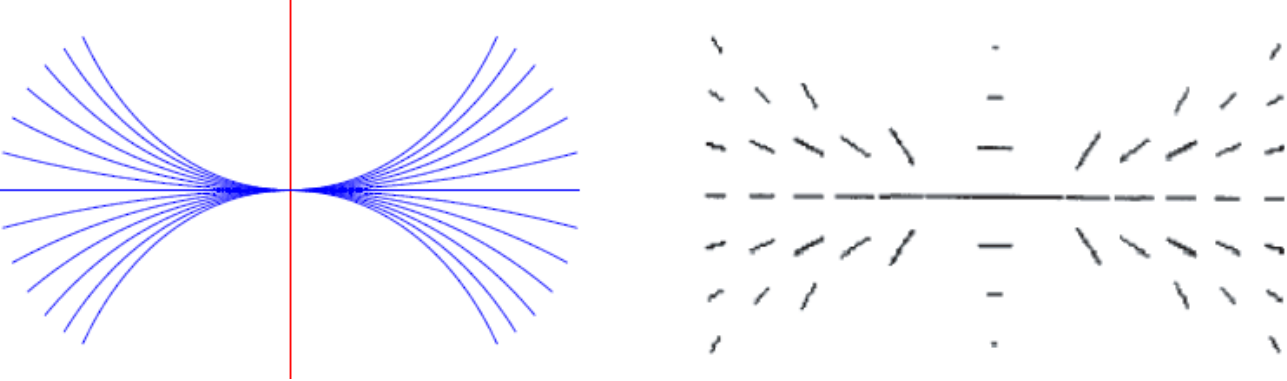}
 	\caption{Top left: integral curves of the vector fields $X_1$ and $X_2$ (blue). Top right: integral curves of the vector fields $X_3$ and $X_4$ (red). Bottom left: their 2D projection. Bottom right: this projection has patters compatible with the measured connectivity patterns (from \cite{Shih-Cheng}).}
\label{fig:curves}
\end{figure} 
A geometrical structure compatible with the observed connectivity is the Riemannian metric $g_F$ which makes the vector fields $X_1, X_2, X_3, X_4$ orthonormal. Indeed, due to the different scale factor, 
$X_1/\sigma$, $X_2/\sigma$ have  a non zero limit, when $\sigma$ goes to $0$, while 
$X_3/\sigma$, $X_4/\sigma$ tend to 0. For this reason, 
this metric couples naturally the vector fields $X_1, X_2$ and $X_3, X_4$. The integral curves starting for a fixed point of the first two and the second two couples of vectors give rise to two families of curves (see fig \ref{fig:curves}, top) whose 2D projection reveals the same pattern of co-axial and trans-axial connections measured by \cite{Shih-Cheng}, validating the model at a neuro-physiological level (see \ref{fig:curves}, bottom). Therefore a decreasing function of the distance can be considered a good model for the connectivity kernel. 

\subsubsection{Non-maxima suppression}
The cortex is equipped with an intracortical neural circuitry which acts within a single hypercolumn. In presence of a visual stimulus, at a point  $x=(x_1, x_2)$, the whole hypercolumn over that point fires, but mechanism of non-maximal
suppression acts, suppressing the output of cells that within the same hypercolumn are not maximal. In this way the connection is able to sharpen the tuning of feature selection over each point $x=(x_1, x_2)$. This selection defines a value of the feature $f$ (scale) at every point. The points of maximal response will be denoted from now on as $\bar f(x)$. The output will then have the following expression:
\begin{equation}
\label{nonmax}O(x, \bar f(x)) = \max_{f}
O(x, f) \end{equation}

\subsubsection{Long range connectivity}
If a family of cells has been described via $R^2 \times F$, with a metric $g_F$ and a connectivity kernel $K_F$, the problem is to describe the action of the connectivity which induces the cortical activity. 
It can be described with a mean field equation, following an approach first proposed
by Wilson and Cowan (\cite{Wilson1972Excitatory}), Amari (\cite{Amari1972Characteristic})  Ermentraut-Cowan  (\cite{Ermentrout1979Temporal}), Bressloff and Cowan  (\cite{Bressloff2003functional}), to quote a few authors. The equation in its general formulation as the following expression:
\begin{equation}
\frac{\partial}{\partial t}a(x,t) = - a(x,t) + \int K_F(x-x', f-f)\psi\Big( a(x',f', t') + O(x',f') - C  \Big)\, dx'df'
\label{erm-cow}
\end{equation} 
where $\psi$ is a sigmoid, $\alpha$ and $\beta$ suitable constants, $C$ is a normalization factor. The equation can be applied in the lifted space or projected in the 2D space, via the non maxima suppression mechanism. 
The associated stationary equation satisfies
\begin{equation}
a(x,t) = \int K_F(x-x', f-f)\psi\Big( a(x',f', t') + O(x',f') - C  \Big)\, dx'df'
\label{stationary}
\end{equation}

\subsection{A model for GOIs related to the orientation }

In \cite{franceschiello2017mathematical} and \cite{franceschiello2017neuro} two main ideas are developed for explaining orientation related GOIs. The initial stimulus is able to modulate the functional geometry of V1 and the geometry induced by the background of the perceived image can induce a perceptual deformation. 

\subsubsection{The metric modulated by the visual stimulus}

The main idea developed in \cite{franceschiello2017mathematical} and \cite{franceschiello2017neuro} is to modify the model for the functional geometry of V1 provided in \cite{citti2006cortical} and to consider that the image stimulus will modulate the connectivity: the new metric will be expressed as 
$$||a(x,f)|| g_F(x,f).$$
When projected onto the visual space, the modulated connectivity gives rise to a Riemannian metric which is at the origin of the visual space deformation.
In the isotropic case, $g_F(x,f)=Id$ is the identity, and the metric reduces to a single positive real value. In this case non maximal suppression within the hypercolumnar structure is sufficient for explaining the mechanism, hence the metric induced on the 2D plane will be simply computed as:
$$||a(x,f(x))|| Id.$$
If at every point we consider a non isotrotropic metric $g_F$ the projection is obtained by an integration along the fiber $F$ at the point $x$. We refer to \cite{franceschiello2017neuro} where the idea is discussed in detail.

\subsubsection{Retrieving the displacement vector fields}
\label{sec:251}

The mathematical question is how to reconstruct the displacement starting from the strain tensor $\textbf{p}$. We think at the deformation induced by a geometrical optical illusion as an isometry between the $\mathbb{R}^2$ plane equipped with the metric $\textbf{p}$ and the $\mathbb{R}^2$ plane with the Euclidean metric $\textbf{Id}$: \begin{equation*}\Phi: (\mathbb{R}^2, \textbf{p}) \rightarrow  (\mathbb{R}^2, \textbf{Id}).\end{equation*}

In strain theory $\textbf{p}$ is called \textit{right Cauchy-Green tensor} associated to the deformation $\Phi$, (for references see \cite{lubliner2008plasticity,marsden1994mathematical}).
It is clear that it is equivalent to find $\Phi$ or the displacement as a map $$\bar{u}(x_1,x_2) = \Phi(x_1,x_2)  - (x_1,x_2), $$ where $(x_1,x_2) \in \mathbb{R}^2$. We can now express the right Cauchy-Green tensor in terms of displacement $u$.  
For \textit{infinitesimal deformations} of a continuum body, in which the displacement gradient is small ($\Arrowvert \nabla \bar{u}\Arrowvert \ll 1$), it is possible to perform a geometric linearization of strain tensor introduced before, in which the non-linear second order terms are neglected. 
Under this assumption it was proved in \cite{franceschiello2017neuro} that $u$ is a solution of the PDE system:
\begin{eqnarray} \label{laplac}
\left\{\begin{array}{cc} 
\Delta {u}_1 = \frac{\partial}{\partial x_1}p_{11} + 2\frac{\partial}{\partial x_2} p_{12} - \frac{\partial^2}{\partial x_1\partial x_2} u_2 & \,\, \quad \mbox{in}\,\, M \\&\\
\Delta {u}_2 = \, \frac{\partial}{\partial x_2} p_{22} + 2\frac{\partial}{\partial x_1}p_{12} -\frac{\partial^2}{\partial x_1\partial x_2} u_1 &\\
&\\
\frac{\partial}{\partial \vec{n}} {u}_1 = 0 &\,\,\,\quad \mbox{in} \,\, \partial M  \\&\\ \frac{\partial}{\partial \vec{n}} {u}_2 = 0 & \\
\end{array}\right. 
\end{eqnarray}
where $M$ is an open subset of $\mathbb{R}^2$ and $\partial M$ is Lipschitz continuous, with normal defined almost everywhere. Solutions for equation \eqref{laplac} are well defined up to an additive constant, which is recovered imposing $u(0,0)=v(0,0)=0$ for symmetry reasons, where $(0,0)$ is the center of our initial domain $M$.
%
%
%
%
%

\section{The model for scale/size GOIs}
In this section we develop a model for scale type illusory phenomena. Once the connectivity is described, it will be used to define a new strain metric tensor. This enable us to adapt the model presented in section \ref{sec:251} to this new features space and to recover the displacement vector fields induced by size perception. 

\subsection{Scale and size of an object} \label{sec:8:1}

 It is well known that simple cells of V1 are able to select scale of an object, which is approximately the distance from the boundary (see for example  \cite{sarti2008symplectic,petitot2008neurogeometrie}). Strictly related to 
the scale is the size of the object, which represents the spatial dimension of the observed element. The Ebbinghaus and Delboeuf illusions (Figure \ref{ebb1}) are phenomena in which the context induces a misperception of the size of the central target, \cite{kunnapas1955influence}. As a first evaluation of the size of the perceptual units in an image is performed at early stages of the visual process, it is possible to adapt the cortical model to introduce a mechanism of non-maximal suppression able to evaluate the scale within an object. 
Therefore we introduce a metric in the position-size space. Following the intuition that there is a relation between functionality of the cells and shape of connectivity, we assume that the connectivity related to scale and size values, which are real quantities, are isotropic. As a result the connectivity  will decrease with the euclidean distance between the objects of the image. Finally we adapt to this metric the displacement algorithm recalled in section 2.6, in order to model the illusion in Figure \ref{ebb1}. 

\subsection{Scale selection in V1}
\label{dist:sele}

 \begin{figure}
 \centering
 \includegraphics[width = 0.5\linewidth]{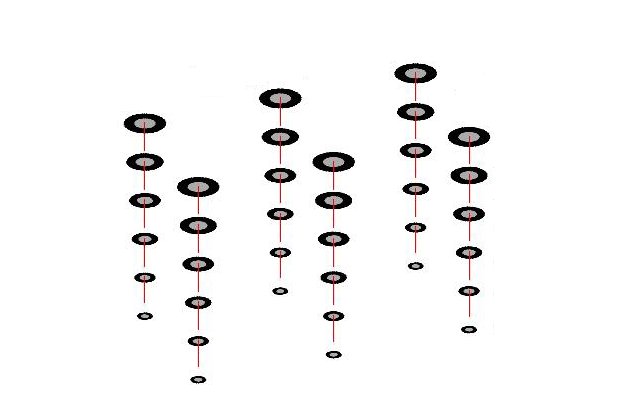}
 \caption{ The hypercolumnar structure for the scale model.}\label{fig:scala}
 \end{figure}

In the previous section we recalled the model of orientation and scale selection of  \cite{sarti2008symplectic} and \cite{petitot2008neurogeometrie}. 
Here we will discard the orientation selection and we focus on the scale detection. This is an isotropic feature, and can be selected by isotropic cells, as for example mexican hat cells, measured by De Angelis (see Figure \ref{fig:2e})
A good model for their receptive profiles are the Laplacian of a Gaussian. 
Denoting $T_{x_1, x_2}$ a translation of a vector $(x_1, x_2)$  and $D_{\sigma}$ the dilation of amplitude $\sigma$, the bank of filters is represented as 
$$ \psi_{x_1, x_2, \sigma}= D_ \sigma T{x_1, x_2}(\psi_0), \text{ where } \psi_0=\Delta G, \text{ and } G(\xi_1, \xi_2) = \frac{1}{\pi } e^{- (\xi_1^2 + \xi_2^2)}.$$ The set of profile is then parametrized by the variables $(x_1, x_2, \sigma),$ where $(x_1, x_2)$ is the spatial position and  $\sigma$ is the scale variable. 
The bank of filters acts on the initial stimulus and the hypercolumns response of simple cells provides an output as the features varies: this mechanism is described in equation \eqref{output_simp_cel}. In the case of having only the scale feature $f=\sigma$ involved, the output reduces to:
\begin{equation}
O(x_1,x_2,\sigma) = \int_M \, I(\xi_1, \xi_2)\, \psi_{(x_1,x_2,\sigma)}\, (\xi_1, \xi_2) \, d\xi_1 d\xi_2 .
\label{output_simp_celts}
\end{equation}
The intra-cortical mechanism selects the maxima over the orientation and scale hypercolumns, providing the selection of two maximal outputs for both features: $\bar{\sigma}$, 
as described in equation 
\eqref{nonmax}: 
\begin{equation}
\label{nonmaxts}O(x, \bar \sigma(x)) = \max_{\sigma}
O(x,   \sigma) \end{equation}

\begin{figure}
\centering
\includegraphics[width = 0.35\linewidth]{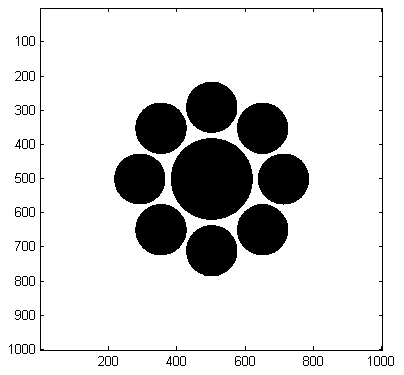} \quad
\includegraphics[width = 0.4\linewidth]{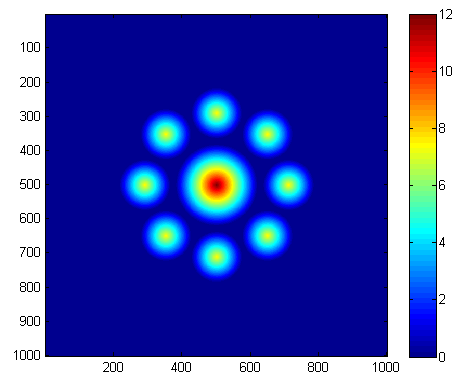}

\caption{Left: the initial stimulus processed. Right: the maximum response $\bar{\sigma}$. For each point the color identifies the distance $\bar{\sigma}$ from the nearest boundary.} \label{barsigma}
\end{figure}

The maximum scale value $\bar\sigma$ represents the distance from the nearest boundary, selected over the hypercolumns containing all the possible distances $\sigma$. This is visualized in figure \ref{fig:scala}, where a bank of filter with different scales, but same orientation, is superimposed to a gray circle (the visual stimulus): the best fit is realized by the central image, whose scale is equal to the distance from the boundary. In figure we visualize an  initial stimulus, the illusion (Figure \ref{barsigma}, left), and apply the scale selectivity maximisation (Figure \ref{barsigma}, right). The level lines of the function $\bar{\sigma}$ are circles, which describe the distance from the boundaries.

\subsection{Size selection} \label{size:sel}
Once the distance function from the boundary $\bar{\sigma}(x_1,x_2)$ has been defined, we assume that 
the action of the connectivity propagates the output within each perceptual unit. 
Since the size $\rho(x_1,x_2)$ of an object can be identified with the maximum distance from the boundary, we postulate the action of a new non maxima suppression procedure, which takes place within the perceptual unit.
This can be implemented through an advection equation
\begin{equation}
\frac{\partial \rho(x_1,x_2)}{\partial t} = | \nabla \bar{\sigma}(x_1,x_2) |
\label{eq:adv}
\end{equation}
This describes a conservation law which associates  a single  size value  $\rho(x_1,x_2)$ to each perceptual unit of the image. 

\begin{figure}
\centering
\includegraphics[width = 0.35\linewidth]{distanzabordo_02} \quad
\includegraphics[width = 0.35\linewidth]{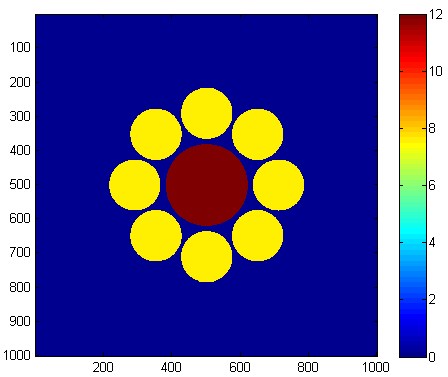}
\caption{Left: representation of $\bar{\sigma}(x_1,x_2)$. Right: propagation of the information within each circle using an advection equation. This allows us to recover for each perceptual unit the corresponding value of size, $\rho(x_1,x_2)$.} \label{adv_eq}
\end{figure}

This step of the algorithm is visualized in Figure \ref{adv_eq}. Starting from the left map representing the value of $\bar{\sigma}(x_1,x_2)$ previously detected, we propagated the maximum distance from the boundary within each circle using an advective equation, see \eqref{eq:adv}.


\subsection{Cortical connectivity for scale type illusion}
Here we introduce the isotropic connectivity accounting for the interaction of points in scale illusions. As we mentioned in section \ref{sec:cc}, we postulate here a strong relation between the functionality of the cells and shape of connectivity. 
Here the set of filters is generated by a fixed one by translation and dilation. 
Since $D_\sigma$ and $T{x_1, x_2}$ commute, then the set $(x_1, x_2, \sigma)$, is a commutative group, and we can consider an isotropic metric on the space. 
Consequently the $g_F$ will simply be the identity. In analogy, we will consider an isotropic metric also in the size space. As a consequence, the connectivity kernel 
will be an exponential decaying function of the Euclidean distance among objects composing the stimulus: 
\begin{equation}
K_F(x-x')  = \exp^{-c\,|x-x'|}
\label{isotropic_kernel}
\end{equation}
The long range spatial interaction decays when the spatial distance of cells increases. Here the kernel is an exponential, but it can be modeled more in general as a function decreasing with the distance. In analogy with the Bressloff- Cowan activity equation recalled in \eqref{stationary}, the stationary activity equation will be expressed as the product between a connectivity kernel and the computed sizes of the objects:
\begin{equation}
a(x) = \int_{\mathbb{R}^2} \exp^{-c\,|x-x'|}(a(x') - \rho_0 )\,\, dx'
\label{equazione_scala}
\end{equation} 
where $\rho_0$ is a global normalization term denoting the \textit{effective size}. It is a mean value for the activity. Since we are interested in evaluating the deformation of the target, we will choose $\rho_0= \rho(0)$, so that $\rho_0$ represents the effective size of the central target. 
 

\subsection{Displacement vector field}

The Euclidean metric, endowing the visual space, will be modulated by stimulus through the cortical activity computed above. Therefore, the metric induced by the stimulus 
in the 2D retinal plane will become:
$$p = a(x_1, x_2) \textbf{Id}$$
at every point. The final step is to adapt the displacement equation to the present setting at every point of the space, and 
according to section \ref{sec:251} we now look for the displacement map 
\begin{equation}u= \Phi - \textbf{Id}, \quad \text{where}
\quad \Phi: (\mathbb{R}^2, \textbf{p}) \rightarrow  (\mathbb{R}^2, \textbf{Id}).\end{equation}
In other word $\Phi$ is the deformation which sends the metric $p$ in the identity at every point. Due to the particular structure of the metric $p$, the equation that in \eqref{laplac} provides the value of $u$ simplifies, as the coefficients $p_{12}$ identically vanish. Indeed in this case the equation for $u$ expresses a Cauchy Riemann - type condition. This means that the solution is harmonic and different from $0$ and results into a radial vector field. 

 \section{Implementation and Results}

In this section the implementation of the presented model is presented and the results discussed. At a first glance, the discrete version of equation \eqref{equazione_scala} becomes:
\begin{equation}
a(x) = \sum_{i=1}^{N} \exp^{-|x-x^i|} (\rho(x^i)- \rho_0 )
\label{discr_eq}
\end{equation}
By simplicity, since $\rho$ and $a$ are locally constant, we assume that $N$ is the number of inducers, and $x^i$, $\forall i$, represents the point in the inducer where the scale is maximal and coincides with the size. Another approximation consists in assuming that the constant $c$ in the exponential map is $1$, meaning
$c=1$. The distance $|x-x'|$ is expressed in pixels, $\rho(x')$ is the size of the inducer at point $x'=(x'_1,x'_2)$. We always consider points of the image in which the maximum of the scale is attained.
The differential problem in \eqref{laplac} is then approximated with a central finite difference scheme and it is solved with a classical PDE linear solver. 
\subsection{Ebbinghaus illusion}

\begin{figure}[hbpt]
\centering
\includegraphics[width = 0.18\linewidth]{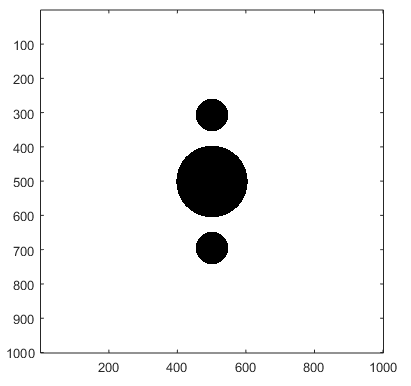} 
\includegraphics[width = 0.18\linewidth]{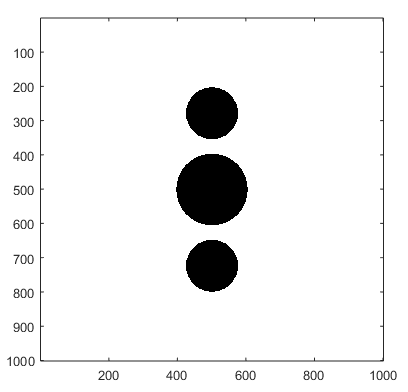} 
\includegraphics[width = 0.18\linewidth]{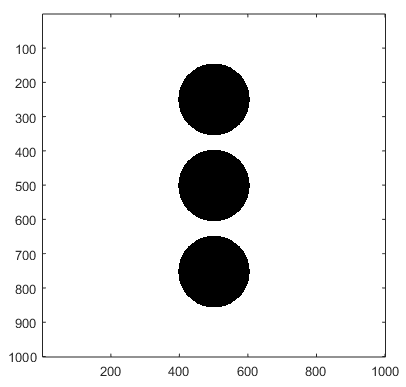} 
\includegraphics[width = 0.18\linewidth]{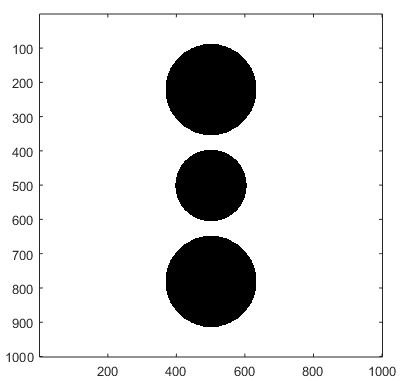} 
\includegraphics[width = 0.18\linewidth]{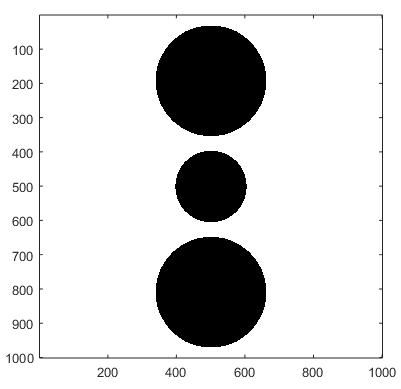} 
\\
\includegraphics[width = 0.18\linewidth]{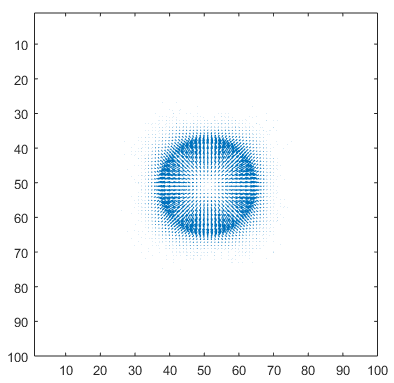} 
\includegraphics[width = 0.18\linewidth]{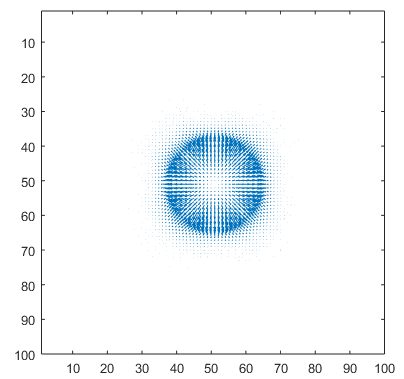} 
\includegraphics[width = 0.18\linewidth]{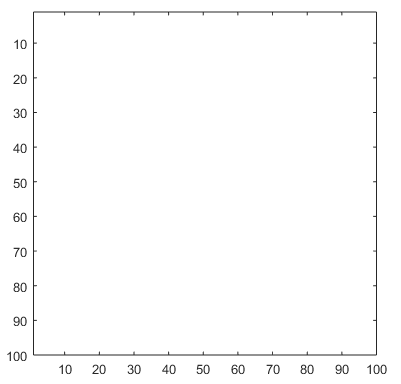} 
\includegraphics[width = 0.18\linewidth]{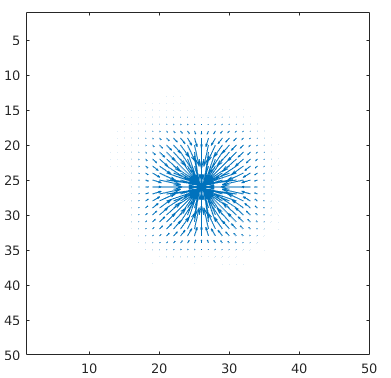} 
\includegraphics[width = 0.18\linewidth]{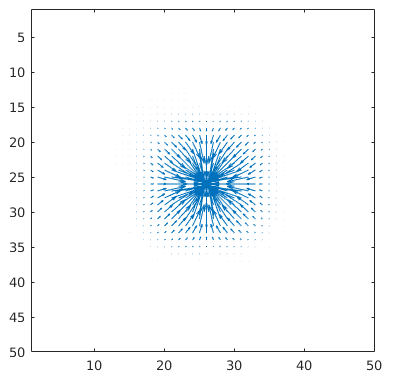} 
\\
\includegraphics[width = 0.18\linewidth]{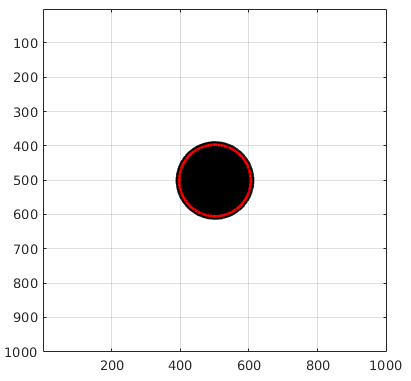} 
\includegraphics[width = 0.18\linewidth]{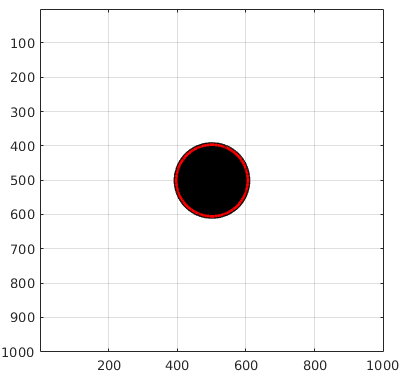} 
\includegraphics[width = 0.18\linewidth]{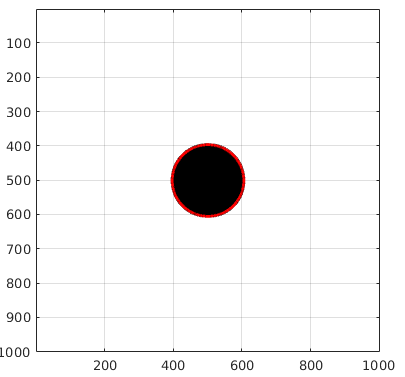} 
\includegraphics[width = 0.18\linewidth]{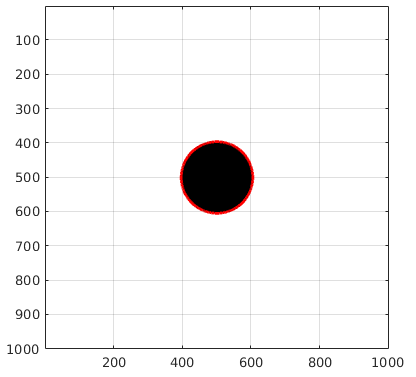}
\includegraphics[width = 0.18\linewidth]{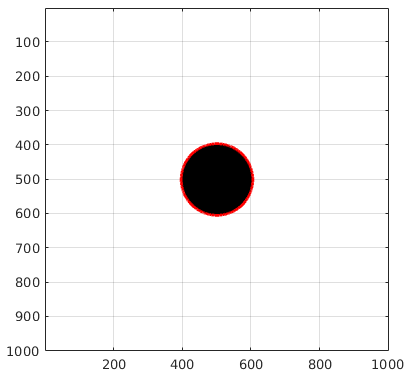} 
\caption{First row: five Ebbinghaus illusion with two inducers of increasing width. Second row: the associated displacement vector fields are visualized.
Third row the deformation of the target is visualized in black (the reference circle is visualized in red). If the inducers are smaller than the target, this expands, if the inducers are larger, the target shrinks.}
\label{fig1_eb}
\end{figure}

We will perform a first test on the Ebbinghaus illusion (Figure \ref{ebb1}, left). This illusion consists in central circle - target - surrounded by a number of circles - inducers. The \textit{perceived size of the target}, which is the perceptual component we want to evaluate in this study varies if the size of the inducers varies \cite{massaro1971judgmental,roberts2005roles} and if the distance between the inducers and the target increases or decreases \cite{roberts2005roles}. 
It has been experimentally reported that whether the inducers are much smaller than the target, this is perceived larger than its actual size. If the dimension of the inducers increases (but remains smaller than the target), the strength of the effect decreases, until the dimension of the target and the inducers are the same. In this last scenario the perceived dimension and the real dimension of the target coincide. If the dimension of the inducers increase, the target is perceived as if it was smaller. This happens independently of the number of the inducers (see fig 9, 10, 11). 
\begin{figure}[htbp]
\centering
\includegraphics[width = 0.18\linewidth]{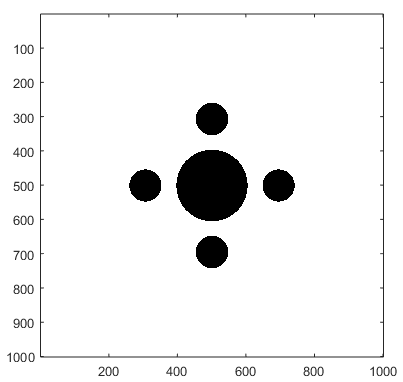} 
\includegraphics[width = 0.18\linewidth]{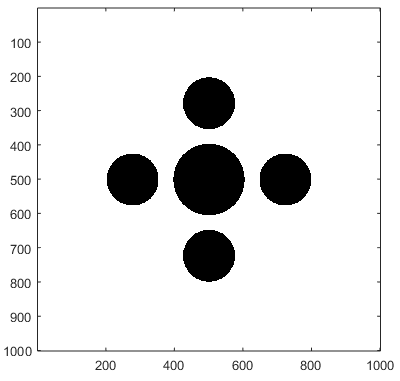}
\includegraphics[width = 0.18\linewidth]{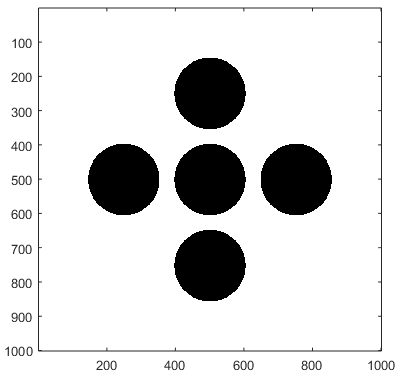}
\includegraphics[width =
0.18\linewidth]{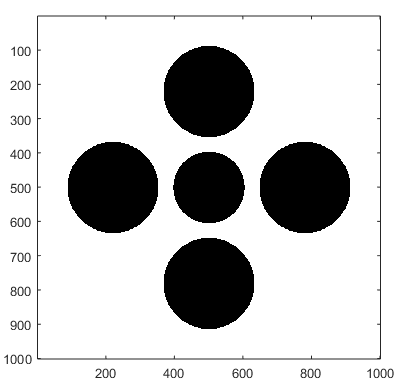}
\includegraphics[width = 0.18\linewidth]{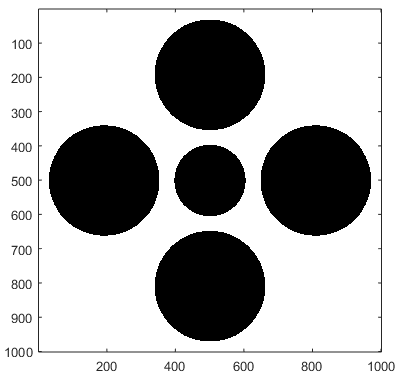}
\\
\includegraphics[width = 0.18\linewidth]{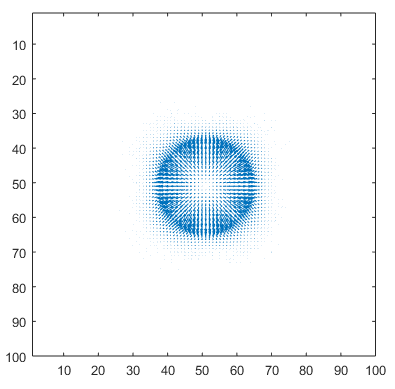} 
\includegraphics[width = 0.18\linewidth]{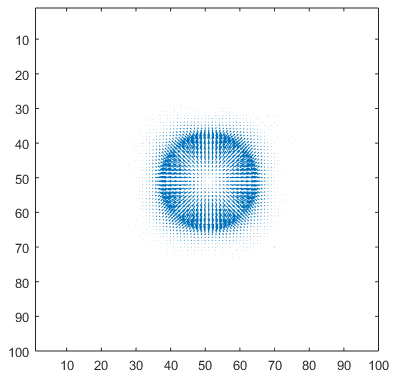} 
\includegraphics[width = 0.18\linewidth]{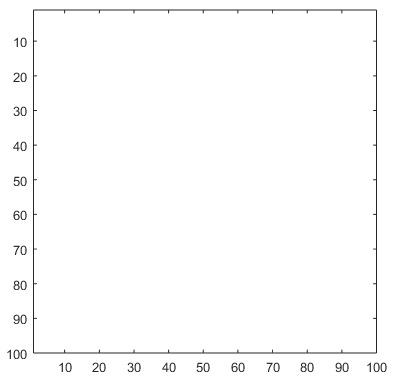} 
\includegraphics[width = 0.18\linewidth]{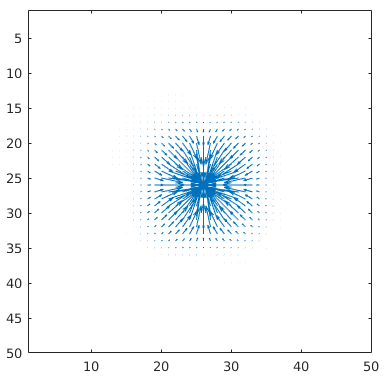} 
\includegraphics[width = 0.18\linewidth]{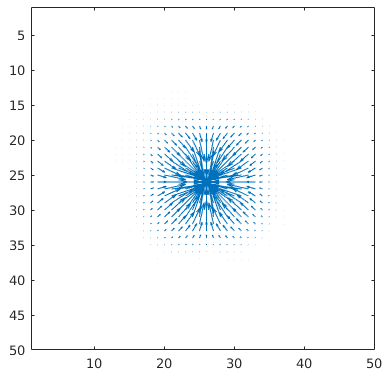} 
\\
\includegraphics[width = 0.18\linewidth]{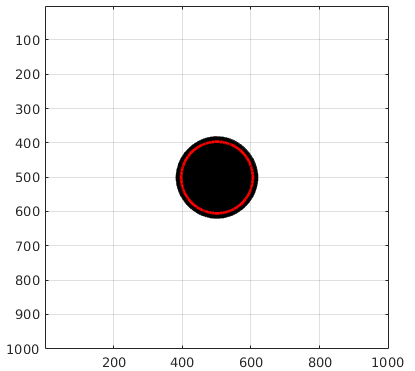} 
\includegraphics[width = 0.18\linewidth]{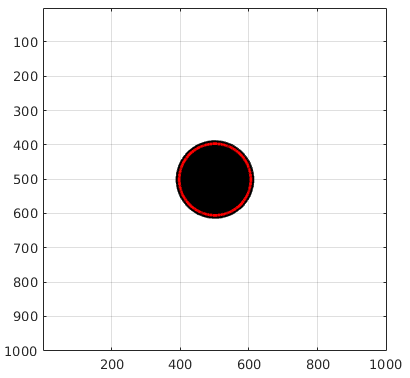} 
\includegraphics[width = 0.18\linewidth]{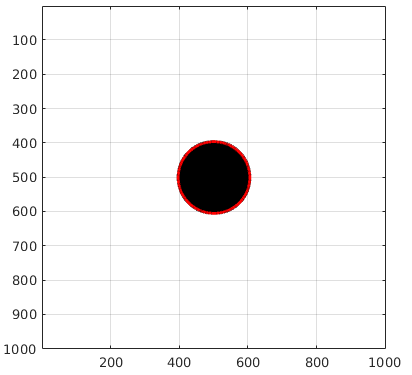} 
\includegraphics[width = 0.18\linewidth]{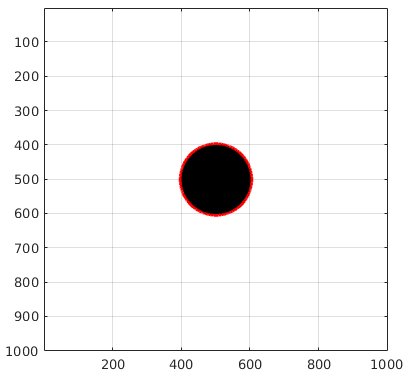} 
\includegraphics[width = 0.18\linewidth]{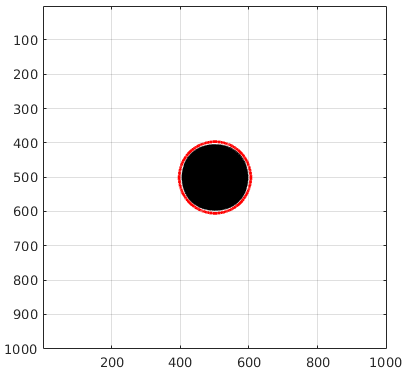} 
\caption{Five Ebbinghaus illusion with four inducers (first row), the corresponding displacement vector field (second row) and the  deformation of the central target (third row). 
}
\label{fig2_eb}
\end{figure}

\subsubsection{Quantitative results: changing the number of inducers}

A quantitative analysis of the phenomenon has been made by Massaro and Anderson \cite{massaro1971judgmental}. In a first experiment, they considered a target circle with diameters of 13 or 17 mm. There were two, four, or six context circles, symmetrically located around the target. The diameters of
the context circles differed from the center circle by 8, 4, 0, - 4, or - 8 mm. for each size of center circle. The distance between proximal edges of center
and context circles was always 6 mm. The figures were presented in four separately
randomized blocks each day, for 2 days, to 10 subjects. They were instructed to judge the apparent size of the target. Responses were made by rotating a wheel that presented single comparison circles. Figure \ref{massano_data1},left summarises the experimental results. In the abscissa the number of surrounding circles is represented, in the ordinate the judged size. If the size of the inducers is fixed, the perceived size of the target grows linearly with the number of inducers.
To validate the model, we started from the same values used in \cite{massaro1971judgmental} as for target measure, number of inducers and distance between target and inducers. 
The diameter of the target wss chosen equal to $\rho_0 = 14.6$, the number of inducers was $N= 2,4,6$ respectively and the size of the inducers was varied as follows: $\rho(x') = \rho_0-8, \rho_0-4, \rho_0, \rho_0+4, \rho_0+ 8$ pixels. Moreover the distance $|x-x'| = 6$ between target and inducers was kept fixed. 
The resulting Ebbinghaus images are depicted on the first row of figures 9, 10, 11. 
In the second row the displacement computed through the infinitesimal strain theory approach is drawn. Finally, the third row contains the perceived central target in black. The red circle is the target reference of the initial stimulus, drawn in order to allow a comparison between the proximal stimulus (displaced image) and the distal one (physical stimulus).
Figure \ref{massano_data1},right, summarises the results found with our model, formally organized as in Massaro and Anderson \cite{massaro1971judgmental}. The modeled and experimental results correctly match. 

\begin{figure}[htbp]
\centering
\includegraphics[width = 0.18\linewidth]{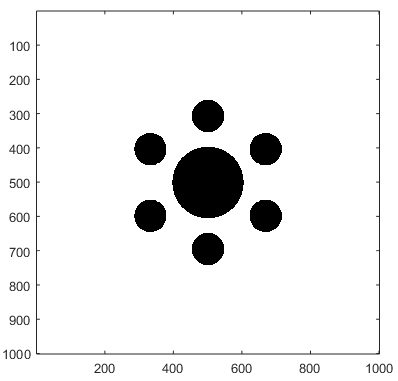} 
\includegraphics[width = 0.18\linewidth]{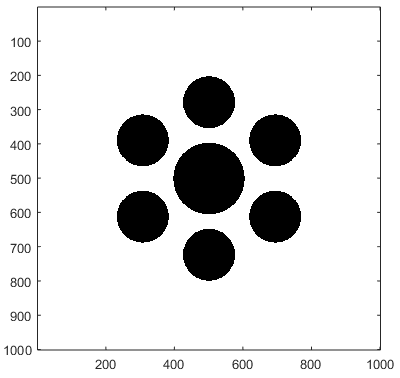} 
\includegraphics[width = 0.18\linewidth]{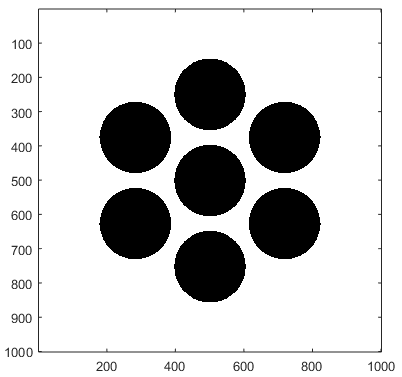}
\includegraphics[width = 0.18\linewidth]{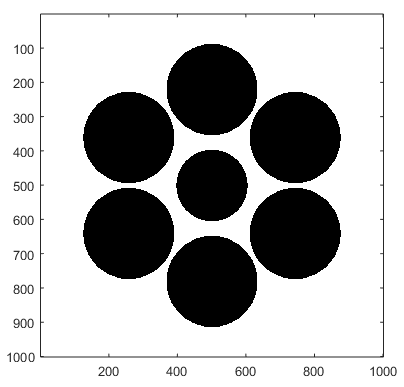} 
\includegraphics[width = 0.18\linewidth]{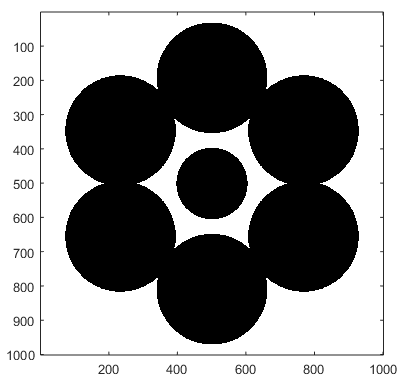} 
\\
\includegraphics[width = 0.18\linewidth]{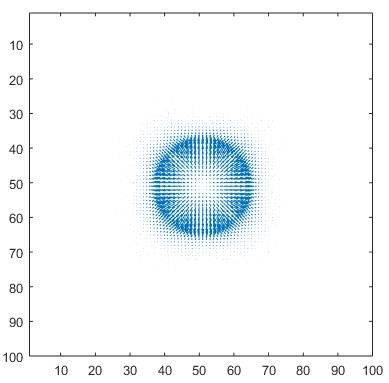} 
\includegraphics[width = 0.18\linewidth]{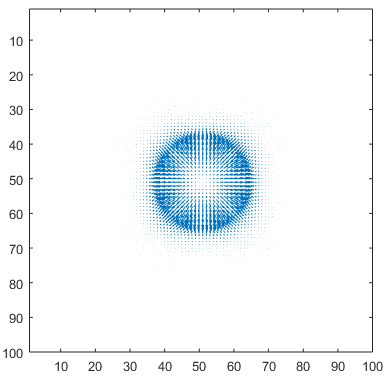} 
\includegraphics[width = 0.18\linewidth]{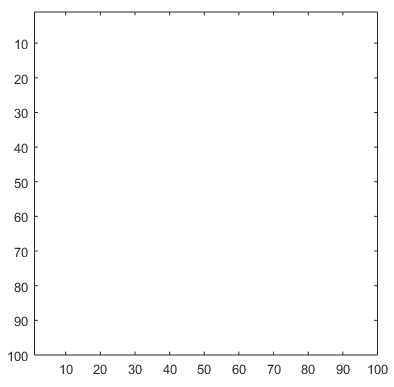}
\includegraphics[width = 0.18\linewidth]{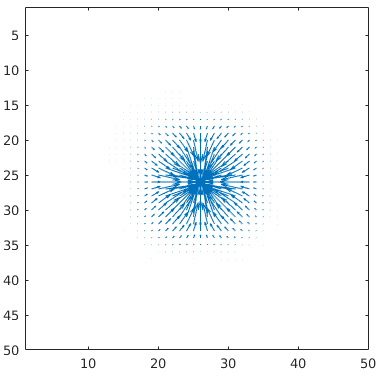} 
\includegraphics[width = 0.18\linewidth]{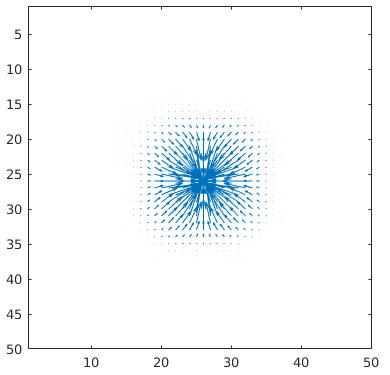} 
\\
\includegraphics[width = 0.18\linewidth]{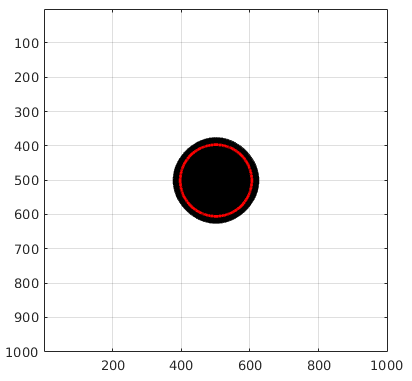}
\includegraphics[width = 0.18\linewidth]{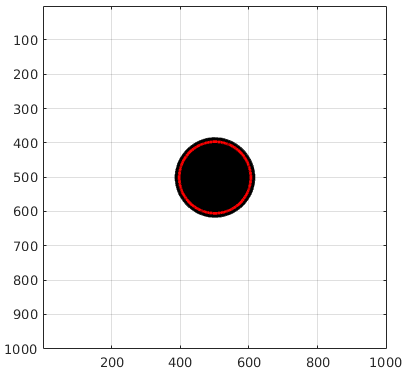}
\includegraphics[width = 0.18\linewidth]{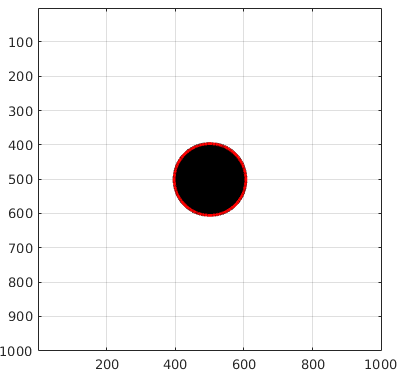} 
\includegraphics[width = 0.18\linewidth]{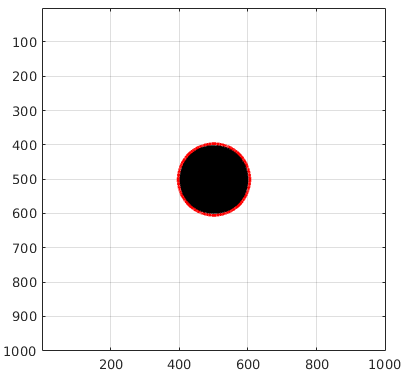} 
\includegraphics[width = 0.18\linewidth]{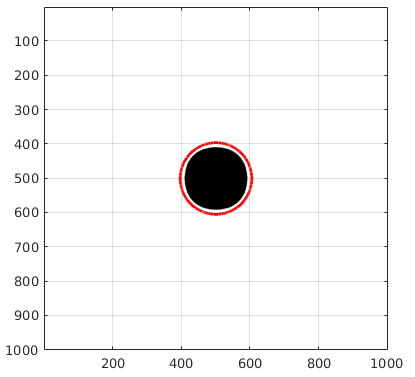} 
\caption{The Ebbinghaus illusion with six inducers, and increasing inducers (first row), the computed  displacement vector field (second row) and the  deformation of the central target (third row).}
\label{fig3_eb}
\end{figure}

\begin{figure}
\centering
\includegraphics[width =0.7 \linewidth]{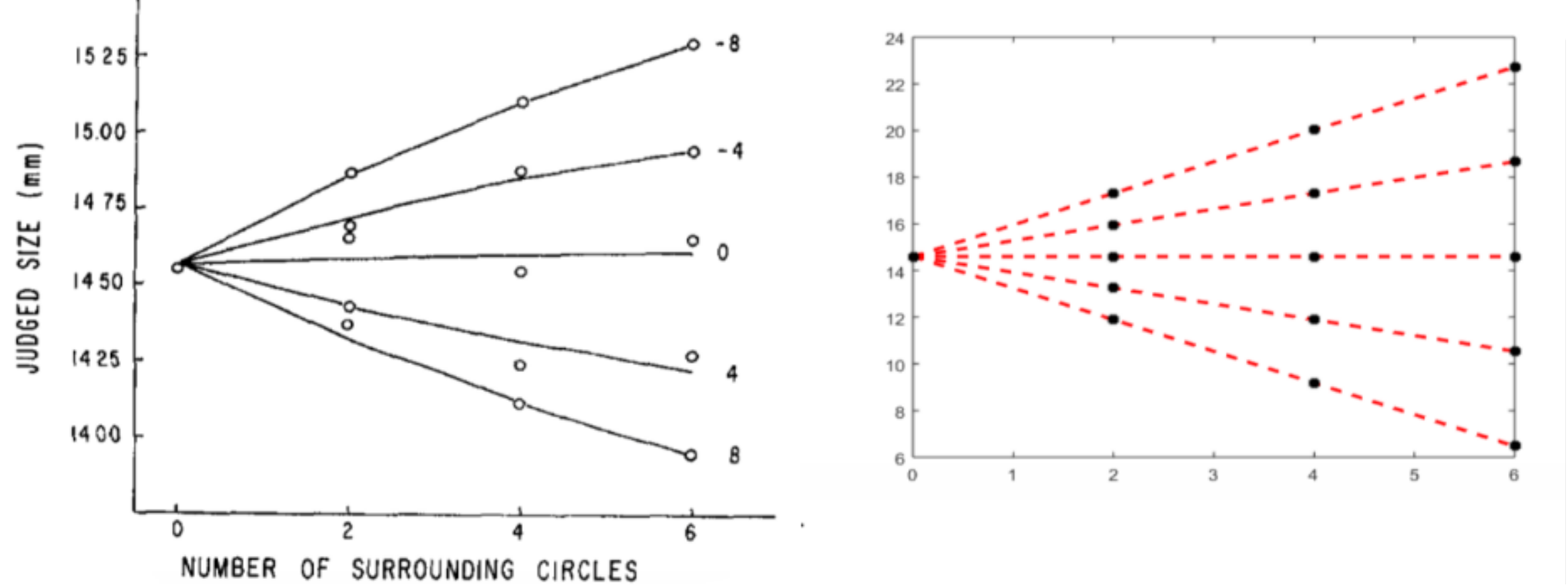}
\caption{Left:  Massaro and Anderson (\cite{massaro1971judgmental}) shown that the perceived size of the central target varies in relationship with the size and the numbers of inducers.  Right: the experimental outcome are exactly reproduced by our model.}
\label{massano_data1}
\end{figure}

\begin{figure}[htbp]
\centering
\includegraphics[width = 0.21\linewidth]{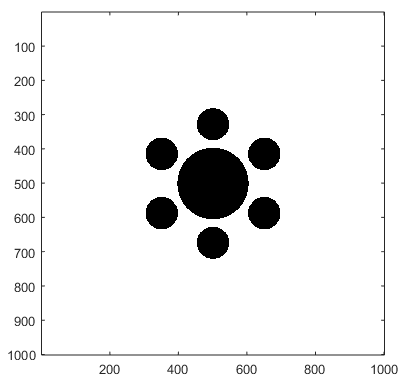}
\includegraphics[width = 0.205\linewidth]{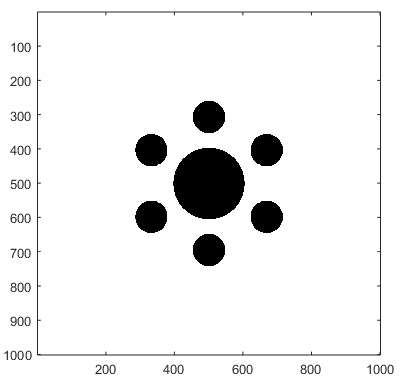}
\includegraphics[width = 0.205\linewidth]{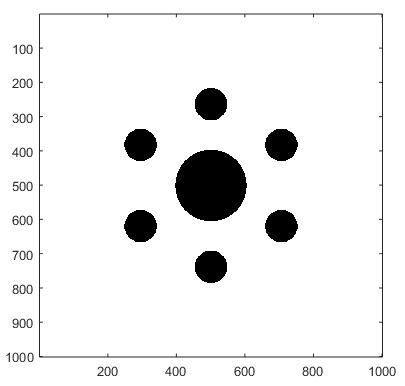}
\includegraphics[width = 0.21\linewidth]{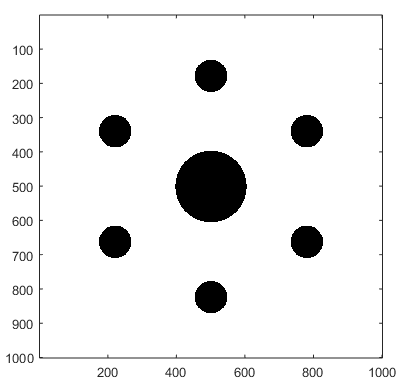}
\includegraphics[width = 0.20\linewidth]{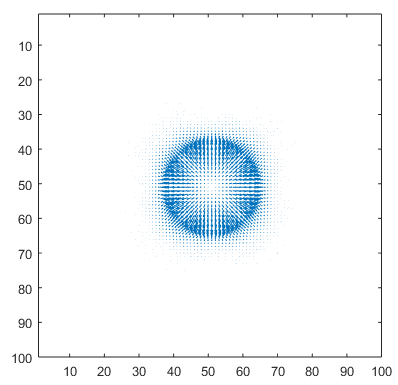} \,
\includegraphics[width = 0.2\linewidth]{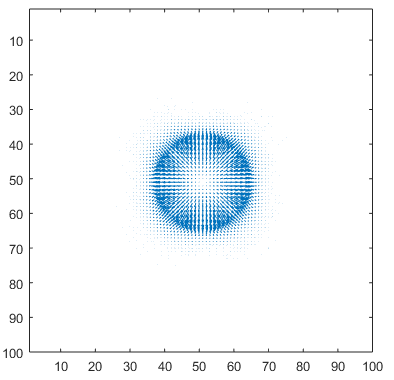} \,
\includegraphics[width = 0.195\linewidth]{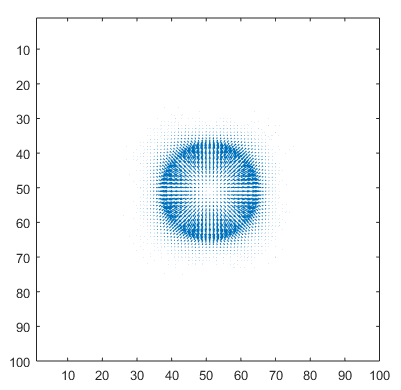} \,
\includegraphics[width = 0.20\linewidth]{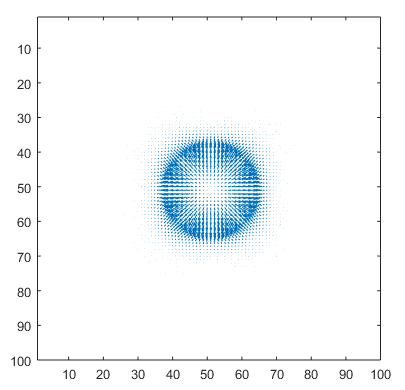} 
\includegraphics[width = 0.21\linewidth]{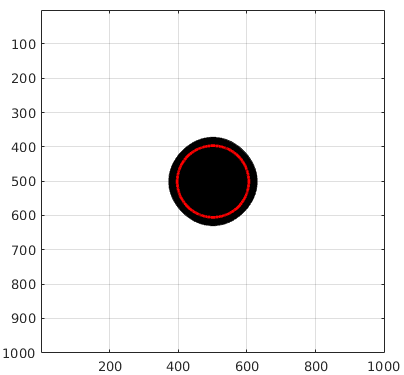}
\includegraphics[width = 0.21\linewidth]{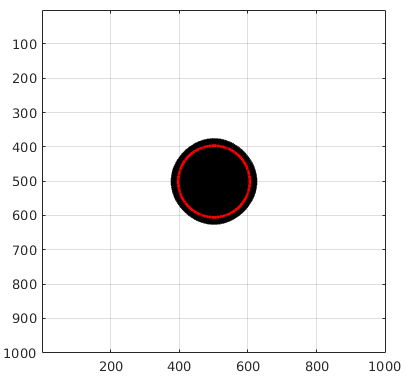}
\includegraphics[width = 0.21\linewidth]{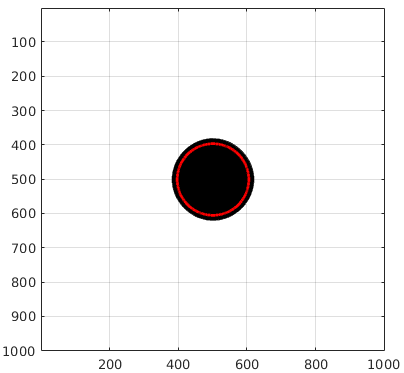} 
\includegraphics[width = 0.21\linewidth]{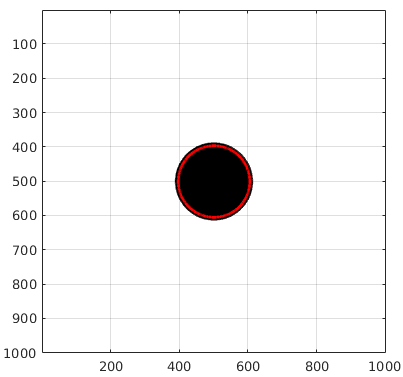} 

\caption{Four Ebbinghaus illusions with increasing distance between the target and the inducers (first row). In the second row the computed displacement vector fields, and in the third row the computed deformation of the central target. If the distance is small, the target expands, while increasing the distance induces a perceptual shrink of the central target, correctly reproduced.}
\label{fig4_eb}
\end{figure}

\begin{figure}
\centering
\includegraphics[width =0.7 \linewidth]{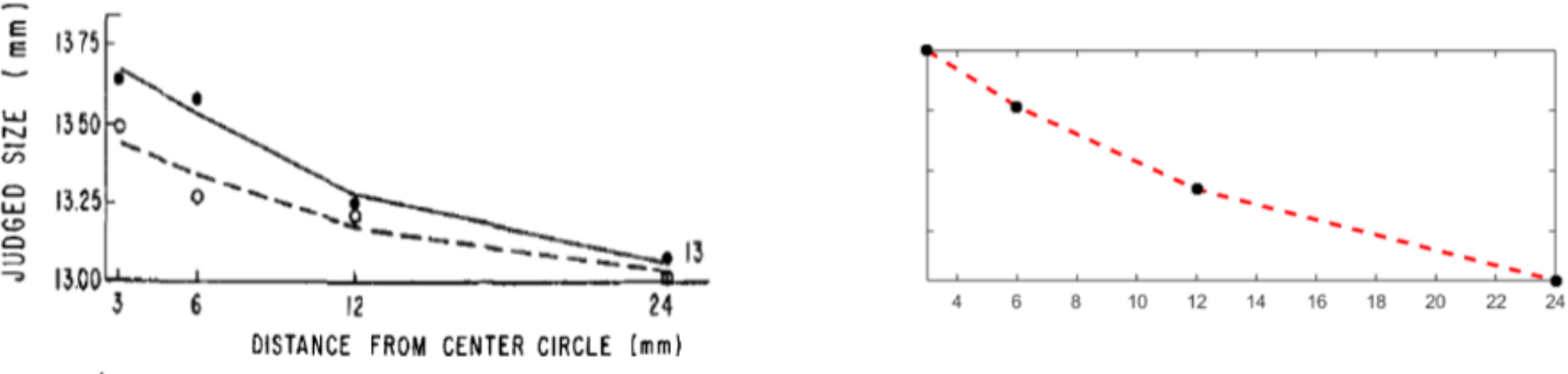}
\caption{Left: Massaro and Anderson (\cite{massaro1971judgmental}) shown how the perceived size of the central target decreases with the distance between the inducers and the target in the Ebbinghaus illusion. Right: we reproduce the experimental results.}
\label{massano_data}
\end{figure}

\subsubsection{Quantitative results: changing the distance between target and inducers}

In a second experiment Massaro and Anderson \cite{massaro1971judgmental} considered a family of Ebbinghaus illusion with six context circles. The diameter of the center circle was 13 mm, the diameter of the context circles was 5 mm. The distances between the proximal edges of center and context circles were 3, 6, 12, and 24 mm respectively. The stimuli were presented to 24 subjects six times in successive randomized blocks. The results are collected in \ref{massano_data}: the distance from the center circle is represented in the x-axis, and the judged size decrease non linearly with the distance. 

We repeated the same experiment with our model. It is represented in figure 13. 
In the first row the four Ebbinghaus illusions, in the second row the computed displacement vector fields, and in the third row the modelled deformation of the  central target. Finally in figure \ref{massano_data} right, we graphically represent the computed outcome in order to perform a comparison with the experimental one. It is easy to check that they correctly match, Figure 14. 

\subsection{Delboeuf illusion}

The Delboeuf illusion, consists in a central black circle (the target) surrounded by an annulus, whose presence induces a misperception of the target size (see figure 15, first row). If the annulus is big, the target tends to shrink. If it is small, the target is perceived as expanding.
\begin{figure}[t]
\centering
\includegraphics[width = 0.18\linewidth]{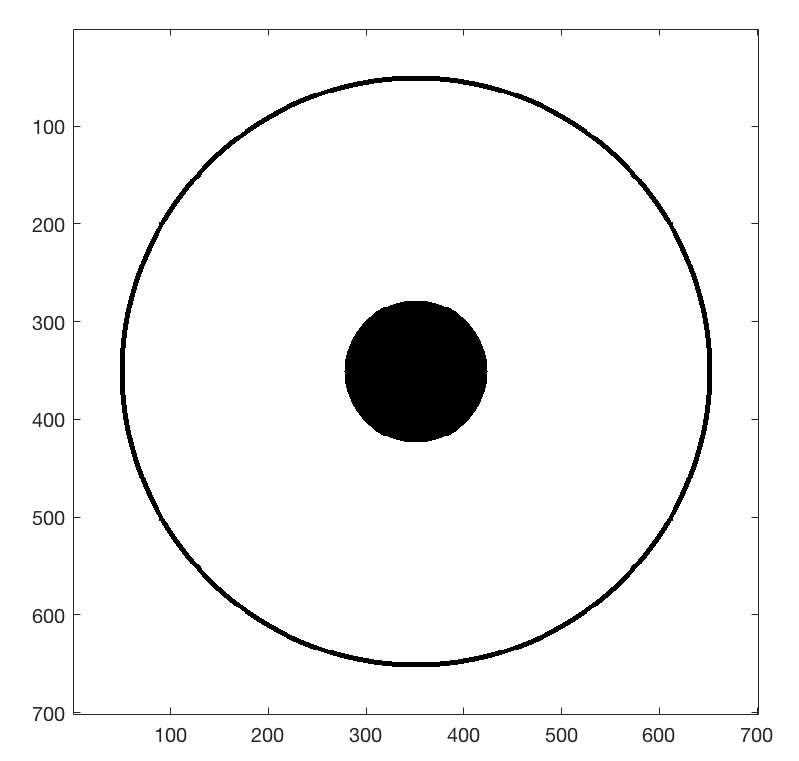} 
\includegraphics[width = 0.18\linewidth]{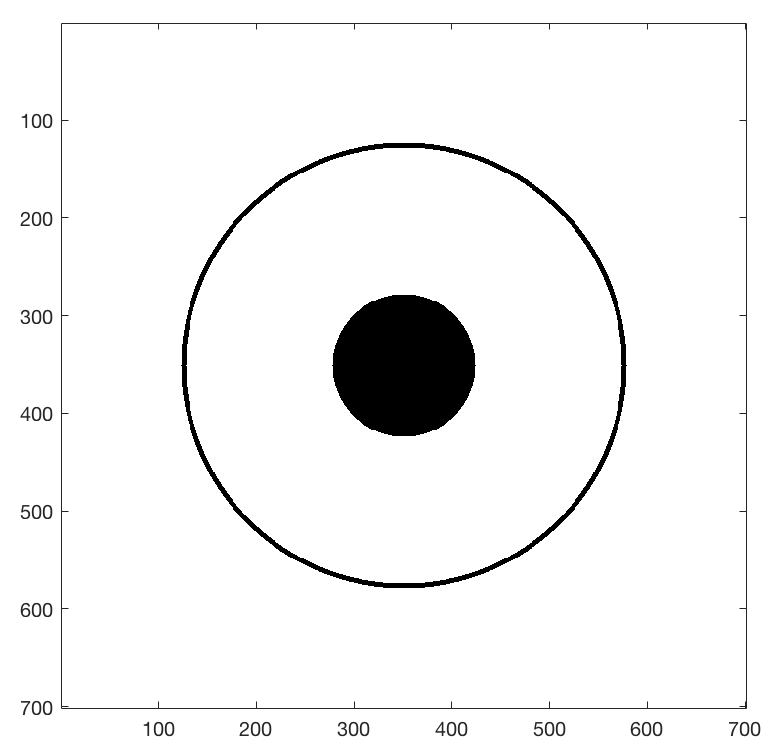} 
\includegraphics[width = 0.18\linewidth]{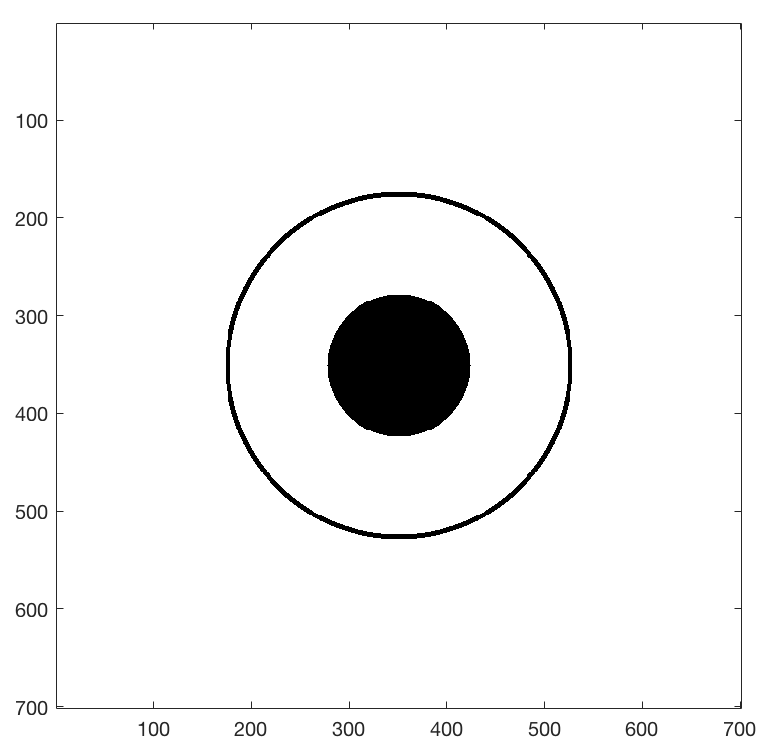} 
\includegraphics[width = 0.18\linewidth]{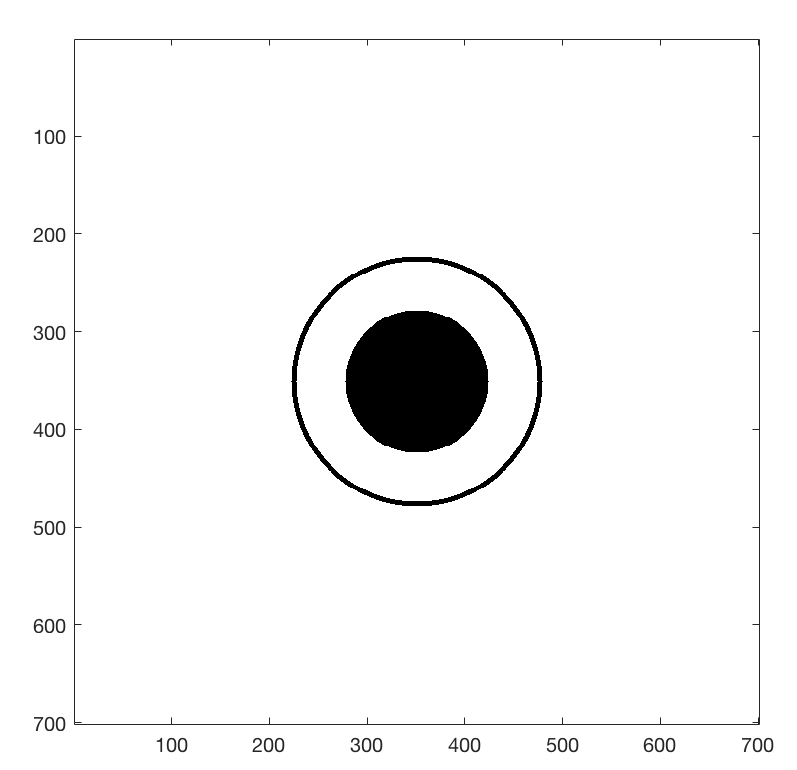} 
\includegraphics[width = 0.18\linewidth]{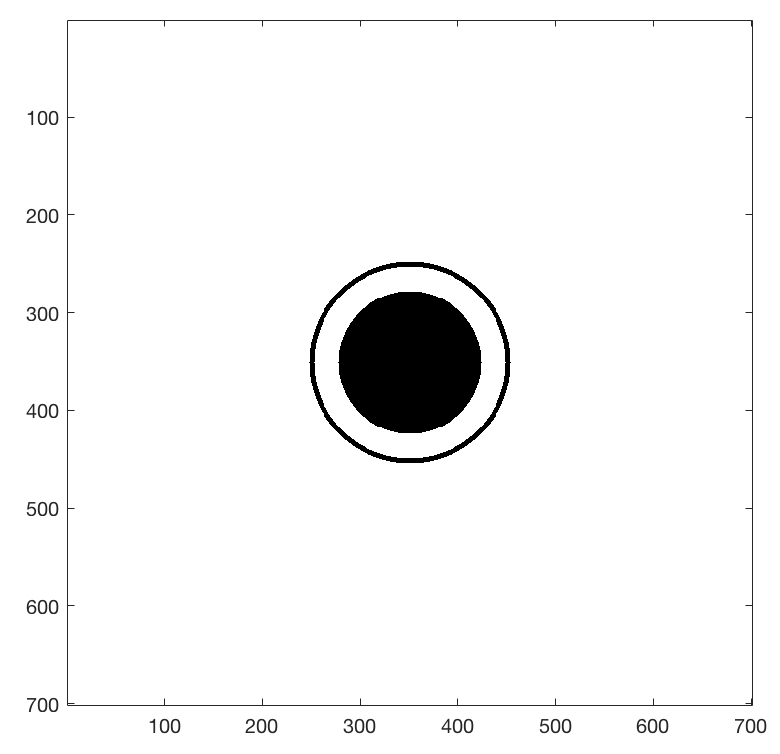} 

\includegraphics[width = 0.18\linewidth]{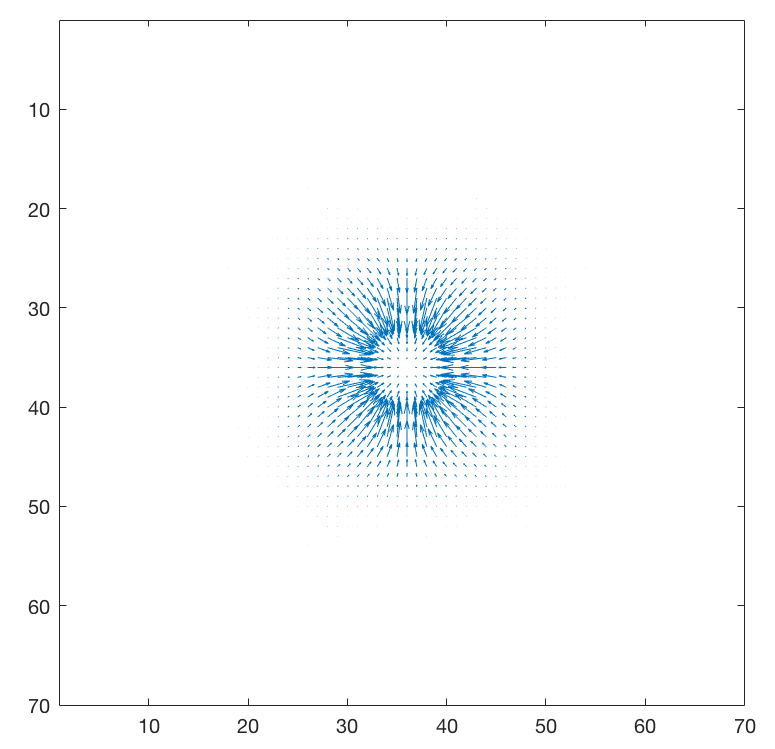} 
\includegraphics[width = 0.18\linewidth]{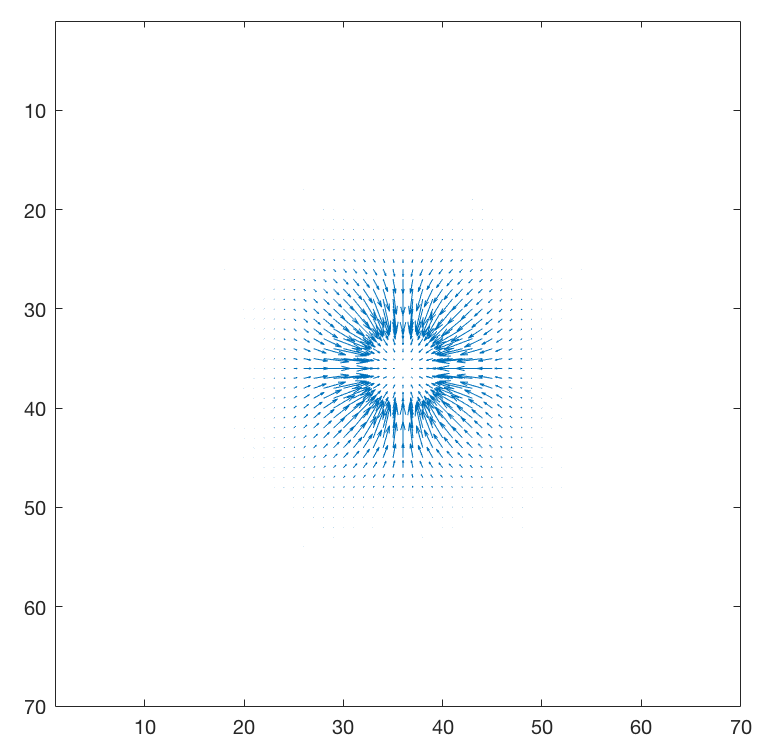} 
\includegraphics[width = 0.18\linewidth]{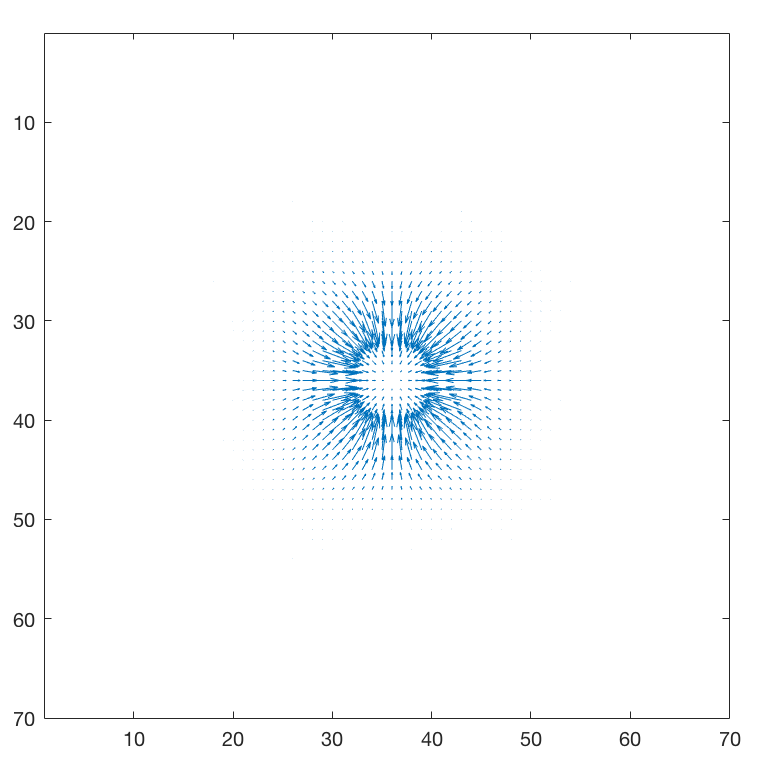} 
\includegraphics[width = 0.18\linewidth]{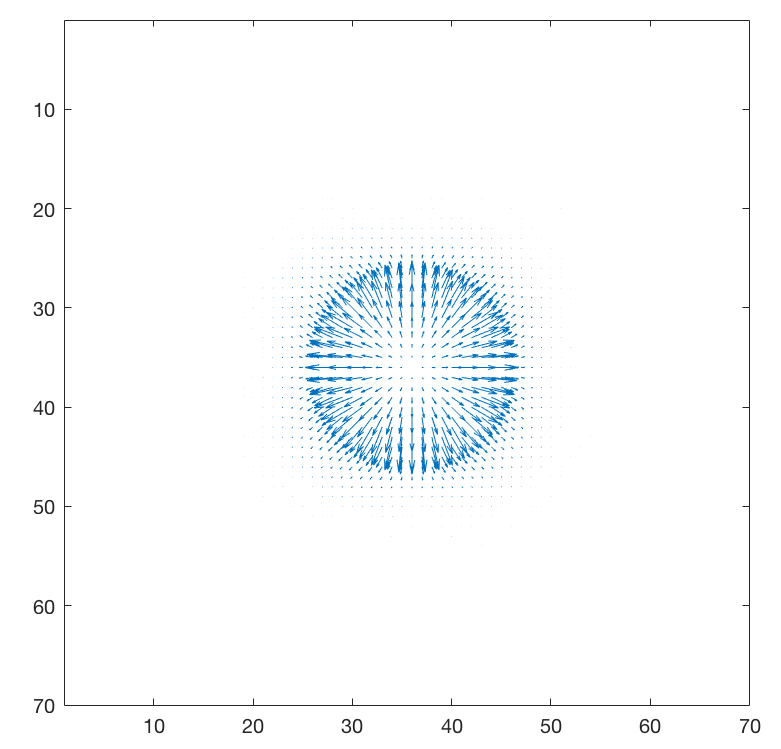} 
\includegraphics[width = 0.18\linewidth]{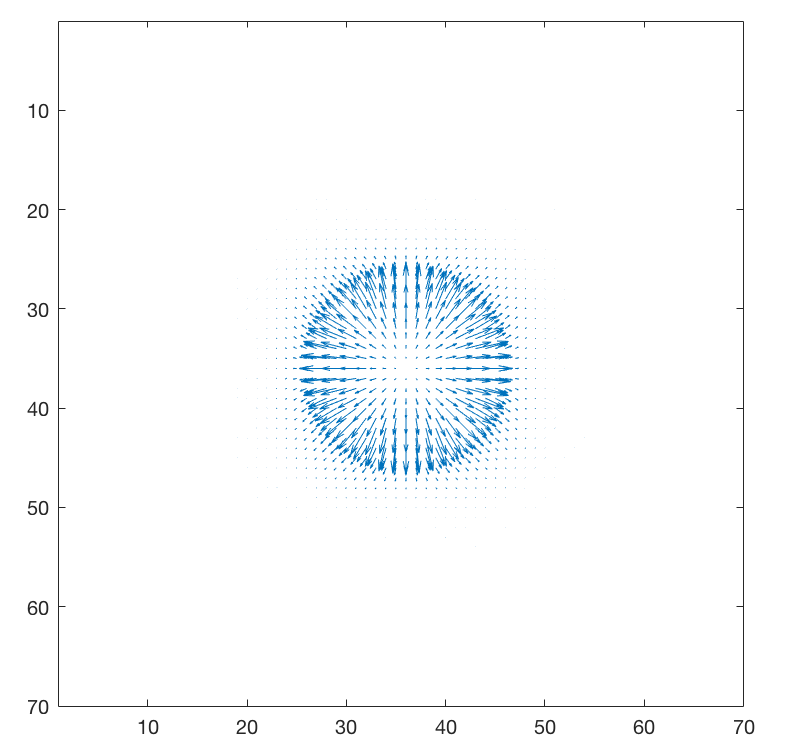} 

\includegraphics[width = 0.18\linewidth]{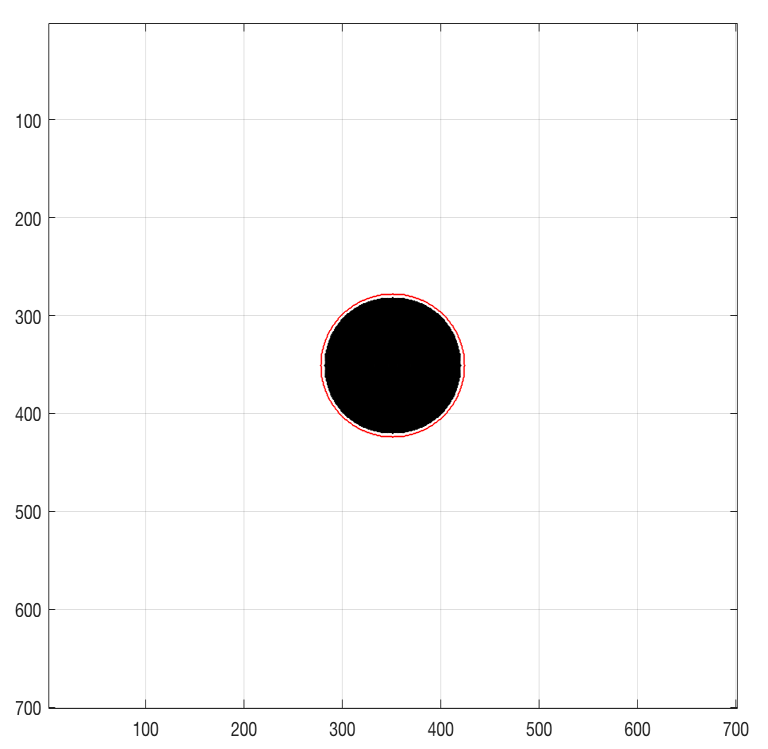} 
\includegraphics[width = 0.18\linewidth]{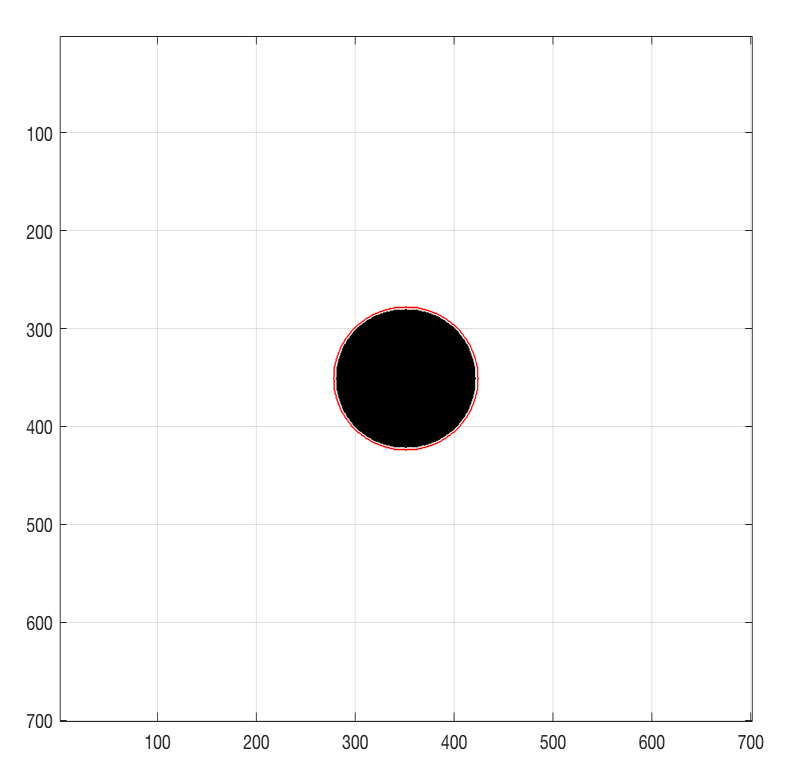} 
\includegraphics[width = 0.18\linewidth]{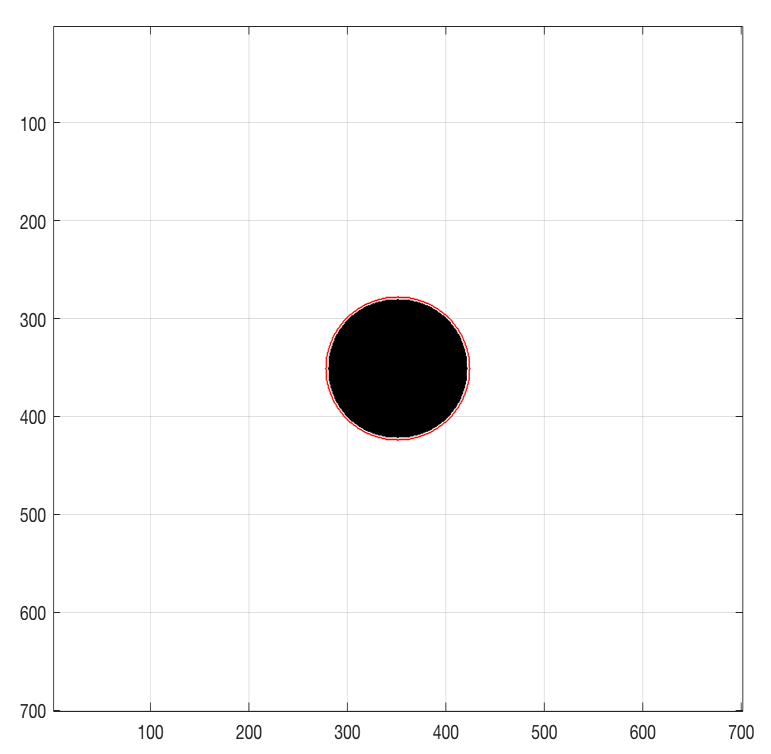} 
\includegraphics[width = 0.18\linewidth]{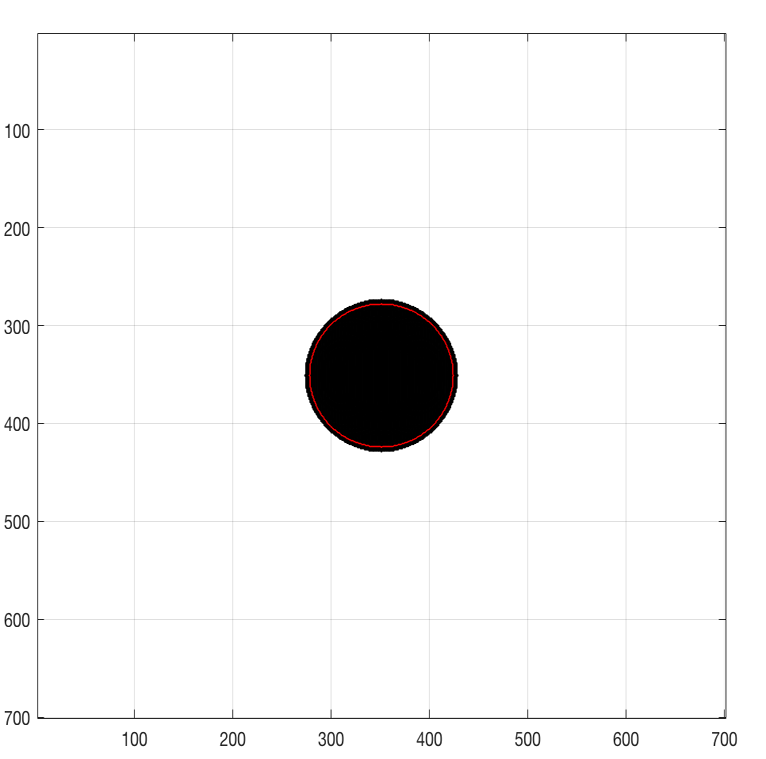}
\includegraphics[width = 0.18\linewidth]{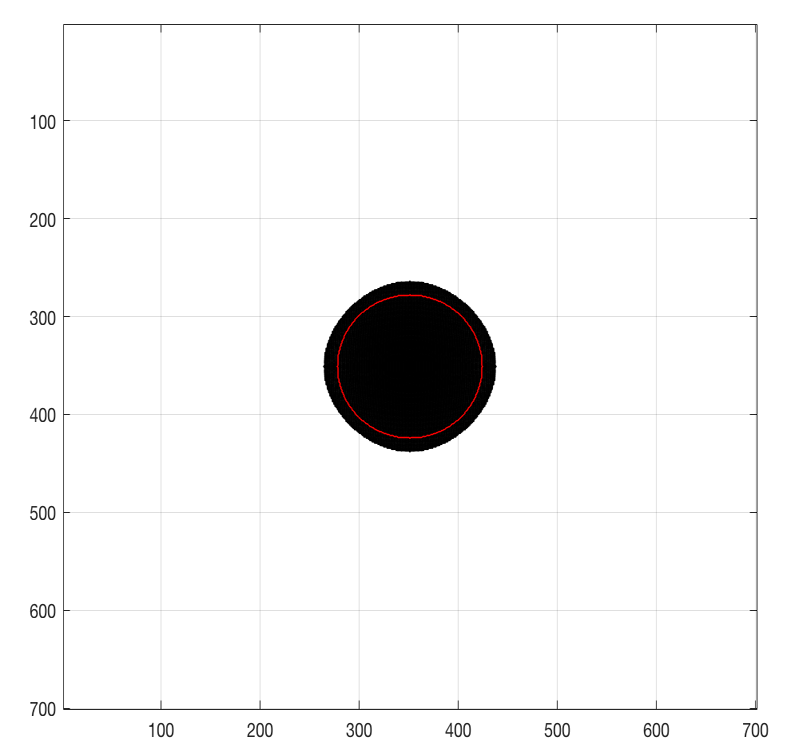} 
\caption{
First row:  five Delbouef illusion with decreasing width for the anulus from left to right.  Second row: the associated displacement vector fields are visualized.  Third row the deformation of the target is visualized in black (the reference circle is visualized in red). 
} \label{delb_img1}
\end{figure}

We apply the presented model to this illusion. Formula \eqref{discr_eq} becomes in this setting:
\begin{equation}
\rho(x) = \exp^{-|x-x'|} (\rho(x')- \rho_0 )
\label{deformation_estimated}
\end{equation}
where $N$ of formula \eqref{discr_eq} is equal to 1, because the inducer is the annulus, $c=1$, the distance $|x-x'|$ is the distance between the center of the target $x$ and the center of the annulus $x'$, expressed in pixels; $\rho(x')$ is the size of the annulus and $\rho_0$ refers again to the effective size. Then $(\rho(x') - \rho_0 )$ expresses the difference between the considered sizes. 

\subsubsection{Discussion of the results}

\begin{figure*}
\centering
\includegraphics[width = 0.43\linewidth]{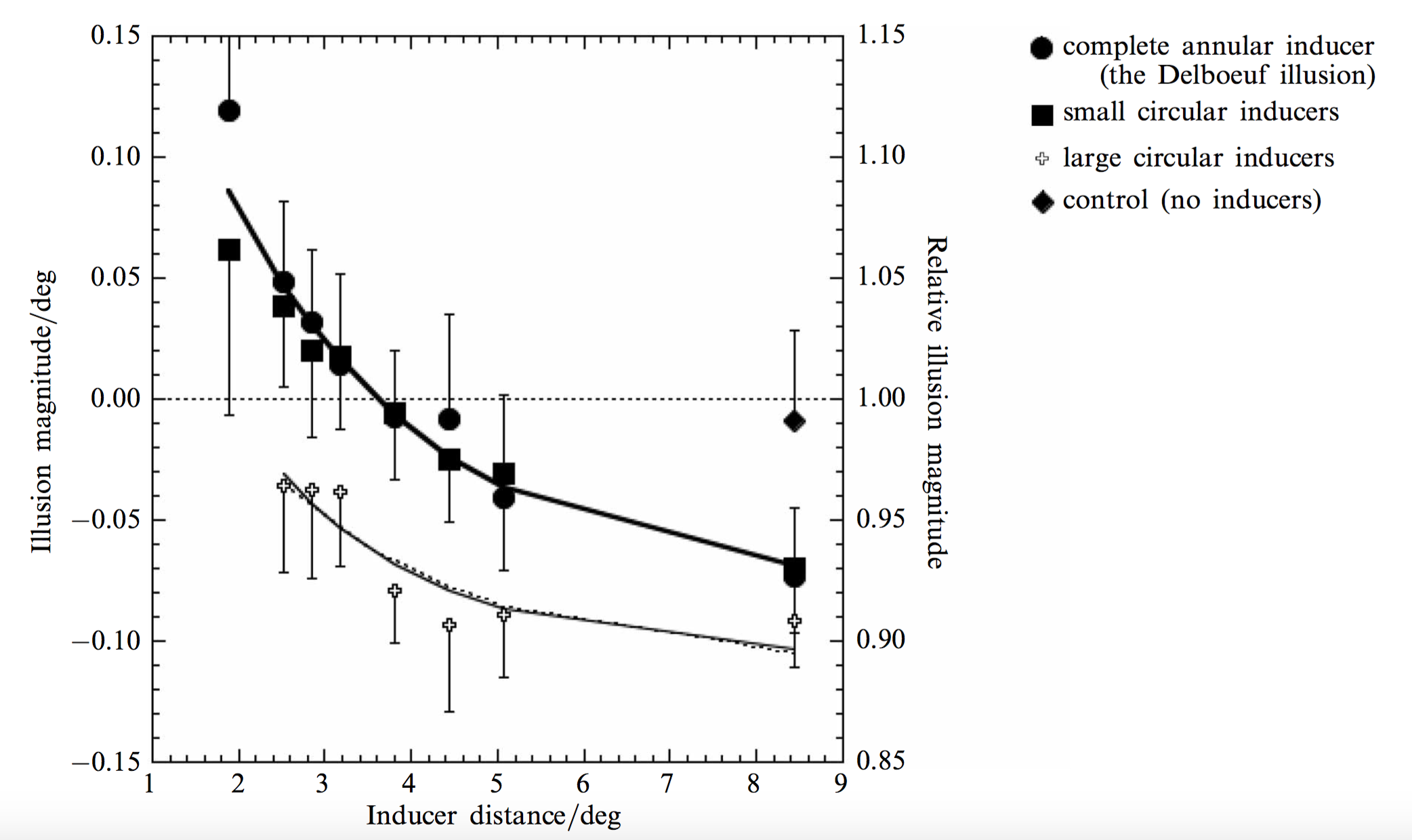}
\includegraphics[width = 0.3\linewidth]{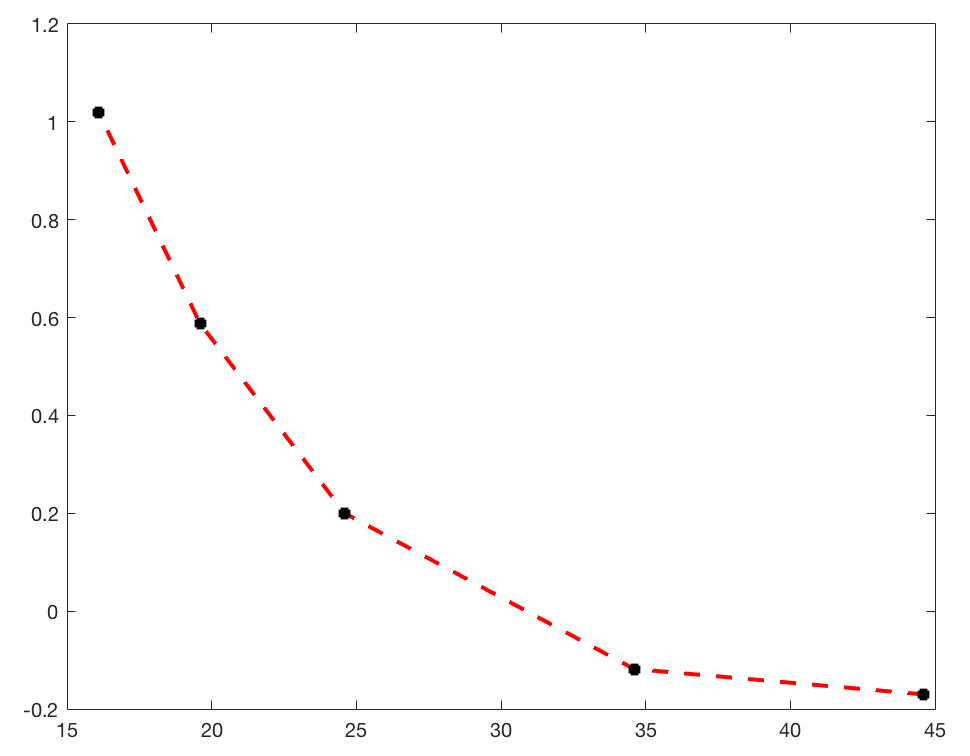} 
\caption{Left: experimental analysis of the decay of the illusion magnitude as a function of the distance between the target and the inducers, from \cite{roberts2005roles}. The circles refers to the Delboeuf illusion. 
Right: the graph shows how the perceived displacement decreases as a function of the distance in our simulations. In the abscissa we put the distance $|x'-x|$ and in the ordinate the computed displacement. 
Our results are in agreement with the ones shown in this experiment.}
\label{disc_delb}
\end{figure*}

We validate the model, comparing with the experimental results of 
(\cite{roberts2005roles}). In this experiment, the authors focused in detail on the effect of inducers and distance. Target size was fixed at 1.27 deg (visual angle). The reported thickness for the anulus is of 0.63 deg., and it was  presented at
distances of 1.90, 2.53, 2.85, 3.17, 3.80, 4.44, 5.07, or 8.45 deg (visual angle). 
Four subjects were asked to evaluate the illusion magnitude, and the experiment was repeated twice. The results are presented in figure \ref{disc_delb}, (left):  the black dots refer to the Delboeuf illusion.

We repeated an analogous simulation applying the presented model. In figure \ref{delb_img1} (first row) we considered five stimuli with a decreasing size of the annulus $\rho(x')$. In the second row we display the computed displacement through the strain theory approach introduced in section 3.5. Finally, the third row contains the reconstructed central target, through formula \eqref{deformation_estimated}. The red circle is the target reference of the initial stimulus, drawn in order to allow a comparison. 
The distance between the center of the annulus and the center of the target $|x'-x|$ decreases from left to right and represents a quantity strictly related with the size of the annulus $\rho(x')$. In fact, increasing $\rho_1$ means we increase the distance between the target and the circumference. 
However the magnitude of the illusion does not depend on the distance alone: when the annulus is big, we perceive a shrinking (see page 454 of \cite{girgus1972interrelationship}), while if the annulus size is decreased, we observe an enlargement of the central target. This variation is explained by an evaluation of the difference in size between the target and the annulus.

The results we obtained are collected in Figure \ref{disc_delb} (left). In the x-axis the distance $|x'-x|$ is reported and along the y-axis the computed displacement. 
The computed displacement decreases as a function of the distance in our computations. The results are in good accordance with the experimental findings of \cite{roberts2005roles}, represented in Figure \ref{disc_delb} (left).

\section{Conclusions}
We proposed a quantitative model for scale-GOIs, inspired by the geometry of the  visual cortex, \cite{sarti2009functional} and a model of orientation related illusions \cite{franceschiello2017neuro}.
We provided here a very general formulation of the neurogeometrical model and of the deformation model in terms of retinal position $(x,y)$ and a feature $f$. 
This allows to choose the scale as encoded feature, and consequently to express the model for size-GOIs. The  model is then validated onto the Ebbinghaus and Delboeuf illusions, and further compared with the results contained in \cite{roberts2005roles, massaro1971judgmental}. The Ebbinghaus and Delboeuf illusions have been often studied together, but it was not clear how to correctly identify the main features playing a role in the second one  \cite{koffka2013principles,gibson1960concept}. The role played by the distance between the inducers and the target was well established, but this was not sufficient to explain the phenomenon. In our model, we conjectured that the misperception is induced by the size difference between the target and the annulus. This is compatible with existing experiments, and correctly explains perceptual effect induced by the phenomena. The advantage of expressing the model in great generality consists in the possibility to extend the same ideas to other features. We think that it would be possible to extend the formulation to consider other GOIs, as for example illusions of movement or those involving higher dimensions encoding. 

\bibliographystyle{spmpsci}
\bibliography{biblio_paper}
\end{document}